\documentclass[a4paper,11pt]{article}
\usepackage{jinstpub} 

\usepackage{multirow}%
\usepackage{booktabs}%
\usepackage{subcaption}

\title{\boldmath Classification of Electron and Muon Neutrino Events for the ESS$\nu$SB Near Water Cherenkov Detector using Graph Neural Networks}

\author[a]{J.~Aguilar}
\author[b]{M.~Anastasopoulos}
\author[r]{D.~Barčot}
\author[c]{E.~Baussan}
\author[b]{A.K.~Bhattacharyya}
\author[b]{A.~Bignami}
\author[d,e]{M.~Blennow}
\author[f]{M.~Bogomilov}
\author[b]{B.~Bolling}
\author[c]{E.~Bouquerel}
\author[g]{F.~Bramati}
\author[g]{A.~Branca}
\author[g]{G.~Brunetti}
\author[h,1,2]{A.~Burgman\note{Corresponding authors}\note{Current address: Department of Physics, Stockholm University, 106 91 Stockholm, Sweden}}
\author[a]{I.~Bustinduy}
\author[h]{C.J.~Carlile}
\author[h,1]{J.~Cederkall}
\author[i]{T.~W.~Choi}
\author[d,e]{S.~Choubey}
\author[h,1]{P.~Christiansen}
\author[j,b]{M.~Collins}
\author[g]{E.~Cristaldo Morales}
\author[k]{P.~Cupia\l}
\author[y]{D.~D'Ago}
\author[b]{H.~Danared}
\author[c]{J.~P.~A.~M.~de~Andr\'{e}}
\author[c]{M.~Dracos}
\author[l]{I.~Efthymiopoulos}
\author[i]{T.~Ekel\"{o}f}
\author[b]{M.~Eshraqi}
\author[m]{G.~Fanourakis}
\author[n]{A.~Farricker}
\author[o,p]{E.~Fasoula}
\author[q]{T.~Fukuda}
\author[u]{J.~Garc\'ia-Marcos}
\author[b]{N.~Gazis}
\author[m]{Th.~Geralis}
\author[r]{M.~Ghosh}
\author[s]{A.~Giarnetti}
\author[t,u]{G.~Gokbulut}
\author[v]{C.~Hagner }
\author[r]{L.~Halić}
\author[u]{M.~Hooft}
\author[h,1]{K.~E.~Iversen}
\author[u]{N.~Jachowicz}
\author[b]{M.~Jenssen}
\author[b]{R.~Johansson}
\author[m,o]{I.~Karakoulias}
\author[o]{E.~Kasimi}
\author[t]{A.~Kayis Topaksu}
\author[b]{B.~Kildetoft}
\author[r]{B.~Kliček}
\author[o,p]{K.~Kordas}
\author[w]{A.~Leisos}
\author[b,h,3]{M.~Lindroos\note{deceased}}
\author[x]{A.~Longhin}
\author[b]{C.~Maiano}
\author[g]{S.~Marangoni}
\author[l]{C.~Marrelli}
\author[s]{D.~Meloni}
\author[y]{M.~Mezzetto}
\author[b]{N.~Milas}
\author[a]{J.L.~Muñoz}
\author[u]{K.~Niewczas}
\author[t]{M.~Oglakci}
\author[d,e]{T.~Ohlsson}
\author[i]{M.~Olveg\r{a}rd}
\author[w]{M.~Pari}
\author[h,1,4]{J.~Park\note{Current address: Center for Exotic Nuclear Studies, Institute for Basic Science, 34126 Daejeon, Republic of Korea}}
\author[b]{D.~Patrzalek}
\author[f]{G.~Petkov}
\author[o,p]{Ch.~Petridou}
\author[c]{P.~Poussot}
\author[m]{A.~Psallidas}
\author[y]{F.~Pupilli}
\author[z]{D.~Saiang}
\author[o,p]{D.~Sampsonidis}
\author[g]{A.~Scanu}
\author[c]{C.~Schwab}
\author[a]{F.~Sordo}
\author[aa]{A.~Sosa}
\author[m]{G.~Stavropoulos}
\author[r]{M.~Stipčević}
\author[b]{R.~Tarkeshian}
\author[g]{F.~Terranova}
\author[v]{T.~Tolba}
\author[b]{E.~Trachanas}
\author[f]{R.~Tsenov}
\author[w]{A.~Tsirigotis}
\author[o]{S.~E.~Tzamarias}
\author[u]{M.~Vanderpoorten}
\author[f]{G.~Vankova-Kirilova}
\author[ab]{N.~Vassilopoulos}
\author[d,e]{S.~Vihonen}
\author[c]{J.~Wurtz}
\author[c]{V.~Zeter}
\author[m]{O.~Zormpa}

\affiliation[a]{Consorcio ESS-bilbao, Parque Científico y Tecnológico de Bizkaia, Laida Bidea, Edificio 207-B, 48160 Derio, Bizkaia, Spain}
\affiliation[b]{European Spallation Source, Box 176, SE-221 00 Lund, Sweden}
\affiliation[c]{IPHC, Universit\'{e} de Strasbourg, CNRS/IN2P3, Strasbourg, France}
\affiliation[d]{Department of Physics, School of Engineering Sciences, KTH Royal Institute of Technology, Roslagstullsbacken 21, 106 91 Stockholm, Sweden}
\affiliation[e]{The Oskar Klein Centre, AlbaNova University Center, Roslagstullsbacken 21, 106 91 Stockholm, Sweden}
\affiliation[f]{Sofia University St. Kliment Ohridski, Faculty of Physics, 1164 Sofia, Bulgaria}
\affiliation[g]{University of Milano-Bicocca and INFN Sez. di Milano-Bicocca, 20126 Milano, Italy}
\affiliation[h]{Department of Physics, Lund University, P.O Box 118, 221 00 Lund, Sweden}
\affiliation[i]{Department of Physics and Astronomy, FREIA Division, Uppsala University, P.O. Box 516, 751 20 Uppsala, Sweden}
\affiliation[j]{Faculty of Engineering, Lund University, P.O Box 118, 221 00 Lund, Sweden}
\affiliation[k]{AGH University of Krakow, al. A. Mickiewicza 30, 30-059 Krakow, Poland }
\affiliation[l]{CERN, 1211 Geneva 23, Switzerland}
\affiliation[m]{Institute of Nuclear and Particle Physics, NCSR Demokritos, Neapoleos 27, 15341 Agia Paraskevi, Greece}
\affiliation[n]{Cockroft Institute (A36), Liverpool University, Warrington WA4 4AD, UK}
\affiliation[o]{Department of Physics, Aristotle University of Thessaloniki, Thessaloniki, Greece}
\affiliation[p]{Center for Interdisciplinary Research and Innovation (CIRI-AUTH), Thessaloniki, Greece}
\affiliation[q]{Institute for Advanced Research, Nagoya University, Nagoya 464–8601, Japan}
\affiliation[r]{Center of Excellence for Advanced Materials and Sensing Devices, Ruđer Bo\v{s}kovi\'c Institute, 10000 Zagreb, Croatia}
\affiliation[s]{Dipartimento di Matematica e Fisica, Universit\'a di Roma Tre, Via della Vasca Navale 84, 00146 Rome, Italy}
\affiliation[t]{University of Cukurova, Faculty of Science and Letters, Department of Physics, 01330 Adana, Turkey}
\affiliation[u]{Department of Physics and Astronomy, Ghent University, Proeftuinstraat 86, B-9000 Ghent, Belgium}
\affiliation[v]{Institute for Experimental Physics, Hamburg University, 22761 Hamburg, Germany}
\affiliation[w]{Physics Laboratory, School of Science and Technology, Hellenic Open University, 26335, Patras, Greece }
\affiliation[x]{Department of Physics and Astronomy "G. Galilei", University of Padova and INFN Sezione di Padova, Italy}
\affiliation[y]{INFN Sez. di Padova, Padova, Italy}
\affiliation[z]{Department of Civil, Environmental and Natural Resources Engineering Lule\aa~University~of~Technology, SE-971 87 Lulea, Sweden}
\affiliation[aa]{ISIS, STFC, Rutherford Appleton Laboratory, Harwell Oxford, Didcot OX11 0QX, United Kingdom}
\affiliation[ab]{Institute of High Energy Physics (IHEP) Dongguan Campus, Chinese Academy of Sciences (CAS), Guangdong 523803, China}

\emailAdd{alexander.burgman@fysik.su.se}
\emailAdd{joakim.cederkall@fysik.lu.se}
\emailAdd{peter.christiansen@fysik.lu.se}
\emailAdd{kaare.iversen@fysik.lu.se}
\emailAdd{jcpark@ibs.re.kr}

\abstract{In the effort to obtain a precise measurement of leptonic CP-violation with the ESS$\nu$SB experiment, accurate and fast reconstruction of detector events plays a pivotal role. In this work, we examine the possibility of replacing the currently proposed likelihood-based reconstruction method with an approach based on Graph Neural Networks (GNNs). As the likelihood-based reconstruction method is reasonably accurate but computationally expensive, one of the benefits of a Machine Learning (ML) based method is enabling fast event reconstruction in the detector development phase, allowing for easier investigation of the effects of changes to the detector design. Focusing on classification of flavour and interaction type in muon and electron events and muon- and electron neutrino interaction events, we demonstrate that the GNN reconstructs events with greater accuracy than the likelihood method for events with greater complexity, and with increased speed for all events. Addition<ally, we investigate the key factors impacting reconstruction performance, and demonstrate how separation of events by pion production using another GNN classifier can benefit flavour classification.}


\begin{document}
\maketitle
\flushbottom

\section{Introduction}
\label{sec:intro}

\begin{figure}[!t]
    \centering
    \includegraphics[width=.9\linewidth]{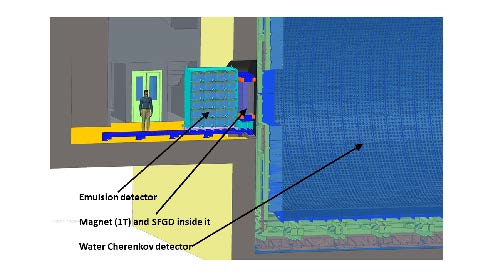}
    \caption{The three detector components of the near detector complex, the emulsion detector, the SFGD and the water Cherenkov detector. The direction of the neutrino beam is left to right, perpendicularly to the face of the WC tank. Illustration from \cite{Alekou_2022}.}
    \label{neardetector}
\end{figure}

The existence of CP violation in the leptonic sector is among the most interesting open questions in particle physics, and can be investigated through the study of neutrino oscillations \cite{Cervera_2000}. The potential impact of the CP violating phase $\delta_{CP}$ on the neutrino oscillation probability is almost 3 times larger at the second oscillation maximum than at the first \cite{Coloma2012}, but requires 9 times more statistics, as the second maximum is located 3 times further away than the first, and the flux is inversely proportional to the square of the distance. This can be achieved by using a more intense neutrino beam (about one order of magnitude), measuring over a longer period (about 9 times longer), or using more efficient event selection algorithms. The European Spallation Source (ESS) linear accelerator (linac) in Lund, Sweden \cite{Garoby_2018} can be modified to provide such a high intensity neutrino beam, and the ESS$\nu$SB is the experiment proposed to measure CP violation using this beam \cite{Alekou_2022, Alekou_2023}. ESS$\nu$SB will measure the flavour composition of the unoscillated neutrino beam immediately after the beam target at the near detector, which consists of a water Cherenkov (WC) detector, a magnetized scintillator detector called the Super Fine-grained Detector (SFGD), and an emulsion-based detector similar to the NINJA detector employed at J-PARC \cite{Fukuda_2017}. The components of the near detector are illustrated in Figure \ref{neardetector}. The flavour composition of the oscillated neutrino beam will be measured at the far detector, composed of two Cherenkov tanks located 340 km from the target such that the flux will contain both the first and second maximum of muon (anti)neutrino to electron (anti)neutrino oscillations \cite{Alekou_2022}. By measuring the flavour composition of both the neutrino and antineutrino beam the oscillation probability can be calculated and $\delta_{CP}$ can be estimated.

To enable these measurements, accurate reconstruction of the neutrino events is a necessity. While accurate reconstruction methods exist and are implemented in neutrino experiments \cite{retro_paper, Missert_2017}, they are often based on likelihood methods and have limits to their performance and reconstruction speed. The latter becomes particularly important in the development stages of the ESS$\nu$SB, as having fast event reconstruction methods enable faster exploration of different detector designs and PMT layouts. Recently, Machine Learning (ML) methods and particularly methods based on Graph Neural Networks (GNNs) have proven useful for neutrino event reconstruction \cite{Abbasi_2022}. In the following sections, we explore the possibilities of replacing the current likelihood-based method with one based on a GNN for classification of flavour and interaction type. 

\section{Methods}

\subsection{Simulation}

The GNN models described in the following sections were trained and subsequently benchmarked against the likelihood method applied to two data sets of Monte Carlo (MC) simulated events (used in the ESS$\nu$SB conceptual design report (CDR) \cite{Alekou_2022} and \cite{Burgman2022}). The events were simulated using the neutrino interaction vertex generator GENIE \cite{Andreopoulos_2010, andreopoulos2015genie} for neutrino interactions, and the water Cherenkov simulation software WCSim \cite{WCSim} for particle transport, and the detector response. WCSim is based on Geant4 \cite{AGOSTINELLI2003250} and developed by the Hyper-Kamiokande collaboration. For these simulations the Near Detector tank is configured to have a volume of 750 m$^3$ with a photomultiplier tube (PMT) coverage of 30\%.

The first dataset contains samples of single charged leptons. These events have been generated homogeneously distributed within the detector tank with an isotropic direction distribution, and uniformly covering kinetic energies from 0 GeV to 1.5 GeV. In the absence of neutrinos and neutrino interactions for this dataset, the flavour of the charged lepton was the target variable for classification. 

The second dataset contains a uniform distribution of interacting neutrino energies from 0 GeV to 1.5 GeV, with pure water as target material. These events were classified by the flavour of the interacting neutrino and later by interaction type. The simulations contain only muon neutrino and electron neutrino events (muons and electrons for the charged lepton simulations), as the tau neutrinos were omitted from this study\footnote{With a beam energy of up to 2 GeV and the tau lepton mass $m_\tau \approx 1.777$ GeV, the tau neutrinos are not produced in significant amounts at the target, and will not partake in CC interaction in the detectors.}. Both events with charged current (CC) and neutral current (NC) interactions were considered.

\subsection{Reconstruction}

\subsubsection{Graph Neural Networks}

The approach developed for this study uses GNNs. GNNs are an ML method based on graph theory, in which data is expressed as graphs of nodes connected by edges, and node information is propagated between the layers of the network through the exchange of information with neighbouring nodes (known as message passing) \cite{GNNs}. In the case of this study, a neutrino interaction event (or an event containing a single charged lepton) is described by a single graph, with each PMT signal being a node with features such as PMT position, signal arrival time and charge recorded. The graph is completed by connecting the nodes with edges according to a K-nearest neighbours scheme (in $x$,$y$,$z$,$t$). 

To train and apply GNN models to the graphs, the GraphNeT \cite{Søgaard2023} framework is used. Initially developed by the IceCube collaboration \cite{Abbasi_2012, Aartsen_2017}, GraphNeT facilitates training of ML models (including GNNs) on data structures and tasks commonly associated with Cherenkov detectors and neutrino telescopes, using the PyTorch \cite{paszke2019pytorchimperativestylehighperformance}, PyTorch Geometric \cite{Fey/Lenssen/2019} and Pytorch Lightning \cite{william_falcon_2020_3828935}, libraries as backend. GraphNeT takes as input tables of PMT signals and event labels, and handles parallel loading of data and building of graphs during training, outputting trained models and predictions for the desired data sets. In addition to useful loss functions and other optimization utilities, GraphNeT contains various model architectures like the DynEdge model. The DynEdge architecture is based on the EdgeConv \cite{wang2019dynamicgraphcnnlearning} message passing scheme, but with the addition of edge reconfiguration between each graph convolution \cite{Abbasi_2022}, and is employed for this study without modifications. 

For each event, the GNN model will output a model score between 0 and 1, representing the probability assigned by the model that the event is of the signal flavour, where a score closer to 1 indicates higher probability of signal flavour. The selection threshold can be adjusted within the this range to obtain a satisfactory sample. For this study, the default signal flavour will be electron flavour, and deviations will be addressed when relevant.

\subsubsection{Likelihood-based Reconstruction}

The baseline method used for classification is based on the fiTQun algoritm \cite{Missert_2017, 10.1093/ptep/ptz015}. A log-likelihood-based reconstruction method developed for the Super-Kamiokande Experiment and based on methods developed for the MiniBooNE experiment \cite{PATTERSON2009206}, it can accurately reconstruct the flavour and kinematic parameters of the interaction products, from which information about their neutrino parents can be inferred. 

For a given event, the fiTQun framework will provide a negative logarithmic likelihood (NLL) estimation for each relevant lepton flavour hypothesis. The model score in this work follows the approach of the CDR \cite{Alekou_2022} and is taken to be the ratio between the NLL of an event being a muon (neutrino) event and the NLL of being an electron (neutrino) event. 

\begin{align}
    \text{Model score} = \frac{ NLL^{\text{fiTQun}_\mu} }{ NLL^{\text{fiTQun}_e} }
\end{align}

\subsubsection{Training}

Both data sets (neutrinos and charged leptons) were classified by neutrino or lepton flavour (and later interaction type and pion production) using the same GNN models, which were trained only on the neutrino event data set. Training  separate models applied to the charged lepton events was investigated, but did not yield any significant increase in performance. The models were trained on 100.000 (67\% training, 33\% validation) events on two NVIDIA Tesla K20 cards and evaluated on $\sim$50.000-300.000 events (depending on the cuts applied and the resulting sample sizes). For a more detailed overview of training and test samples, see Table \ref{tab:samplesizes}. The training was carried out using a linear 1cycle learning rate scheduler \cite{smith2018superconvergence} with early stopping, with all models converging within 50 epochs.

\begin{table}
    \centering
    \begin{tabular}{cccccc}
         & Simulated & After cuts & Training & Test w/  tcuts & Test w/o cuts \\
         \hline
         Charged leptons & 100,000 & 49,325 & - & 49,325 & - \\
         \hline 
         Neutrinos (CC) & 400,000 & 226,425 & 100,000 & 100,000 & 300,000 \\
         \hline
    \end{tabular}
    \caption{Breakdown of the sizes of training and test samples for the charged lepton and CC neutrino interaction data sets. The charged lepton data set was not used for training, and was not tested without cuts. All samples were simulated with a 1:1 ratio of initial lepton/neutrino flavour, and a 1:1 ratio of $\nu$ and $\bar\nu$ for the neutrino sample. The training was divided into 67\% training data and 33\% validation data with the same ratios. \newline
    For the neutrino events, a total of 800,000 events of equal parts CC and NC events were simulated, but as the NC events are disregarded for the majority of this study, they are not covered here.}
    \label{tab:samplesizes}
\end{table}

\subsubsection{Performance Metric}\label{sec:FPR}

The neutrino beam produced at the ESS$\nu$SB target will contain $>$ 98\% muon neutrinos \cite{Alekou_2022} with the remainder dominated by electron-neutrino. The main purpose of the Near Detector (on which this study is focused) is to accurately determine the electron neutrino content of the initial neutrino beam (before oscillations) in order to reduce systematic uncertainties for the oscillation analysis conducted with far detector data. For these reasons, the purity of the electron neutrino sample (which is of greater importance) is more severely affected by the misidentification of muon neutrino events, making it more important to filter out muon neutrino events from the electron neutrino sample than electron neutrinos from the muon neutrino sample. Thus, this work follows the example of the CDR \cite{Alekou_2022} and \cite{Burgman2022} and uses a false positive rate (FPR) of 0.1\% for (muons neutrinos identified as) electron neutrinos and 1\% for (electron neutrinos identified as) muon neutrinos as the target parameter when optimizing the selection criteria. The true positive rate (TPR) is the same as the efficiency, while the FPR is the probability of labelling an event of the background flavour as signal, quantified e.g. for electron neutrinos, as the number of muon neutrino events identified as electron neutrino events relative to the total number of muon neutrinos events, and is thus independent of the ratio between the number of events of each flavour. Purity could also be used for this study, as a pure electron neutrino sample is the key objective, but since we exactly want to measure the beam composition because it is not known precisely and it is not reflected in the simulations, the TPR and FPR are better suited. For illustration, assuming a 90\% TPR (and 98\% $\nu_\mu$, 2\% $\nu_e$), the 0.1\% target TPR will yield an electron neutrino sample of 95\% purity, demonstrated in Eq.\ref{puritycalc}. Extended definitions of true and false positive and negative and their dependence on the signal flavour for flavour classification are illustrated in the confusion matrices in Table \ref{tab:confusion}.

\begin{align}
    \text{n}_{\nu_\mu} = 98\% \cdot 0.1\% = 0.098\% \hspace{1em} 
    \text{n}_{\nu_e} = 2\% \cdot 90\% = 1.8\% \hspace{1em}
    \text{purity}_{\nu_e} = \frac{1.8\%}{1.8\% + 0.098\%} = 95 \%
    \label{puritycalc}
\end{align}

\begin{table}
    \begin{subtable}[c]{0.5\textwidth}
        \centering
        \begin{tabular}{@{}cccc@{}}
            \multicolumn{1}{c}{} &\multicolumn{1}{c}{} &\multicolumn{2}{c}{Prediction} \\ 
            \multicolumn{1}{c}{} & 
            \multicolumn{1}{c}{} & 
            \multicolumn{1}{c}{$\nu_\mu$} & 
            \multicolumn{1}{c}{$\nu_e$} \\ 
            \cmidrule{2-4}
            \multirow[c]{2}{*}{\rotatebox[origin=tr]{90}{Truth}}
            & $\nu_\mu$  & True Positive (TP) & False Negative (FN)   \\[1.5ex]
            & $\nu_e$  & False Positive (FP)   & True Negative (TN) \\ 
            \cmidrule{2-4}
        \end{tabular}
        \subcaption{$\nu_\mu$ signal}
    \end{subtable}
    \hspace{2em}
    \begin{subtable}[c]{0.5\textwidth}
        \centering
        \begin{tabular}{@{}cccc@{}}
            \multicolumn{1}{c}{} &\multicolumn{1}{c}{} &\multicolumn{2}{c}{Prediction} \\ 
            \multicolumn{1}{c}{} & 
            \multicolumn{1}{c}{} & 
            \multicolumn{1}{c}{$\nu_e$} & 
            \multicolumn{1}{c}{$\nu_\mu$} \\ 
            \cmidrule{2-4}
            \multirow[c]{2}{*}{\rotatebox[origin=tr]{90}{Truth}}
            & $\nu_e$  & TP & FN   \\[1.5ex]
            & $\nu_\mu$  & FP   & TN \\ 
            \cmidrule{2-4}
        \end{tabular}
        \subcaption{$\nu_e$ signal}
    \end{subtable}
    \caption{Confusion matrices for flavour classification with the signal flavours $\nu_\mu$ (a) and $\nu_e$ (b).}
    \label{tab:confusion}
\end{table}

\subsection{Likelihood-based Cuts}\label{sec:cuts}

\begin{figure}[!t]
    \centering
    \begin{subfigure}[b]{\textwidth}
        \centering
        \includegraphics[width=0.5\linewidth]{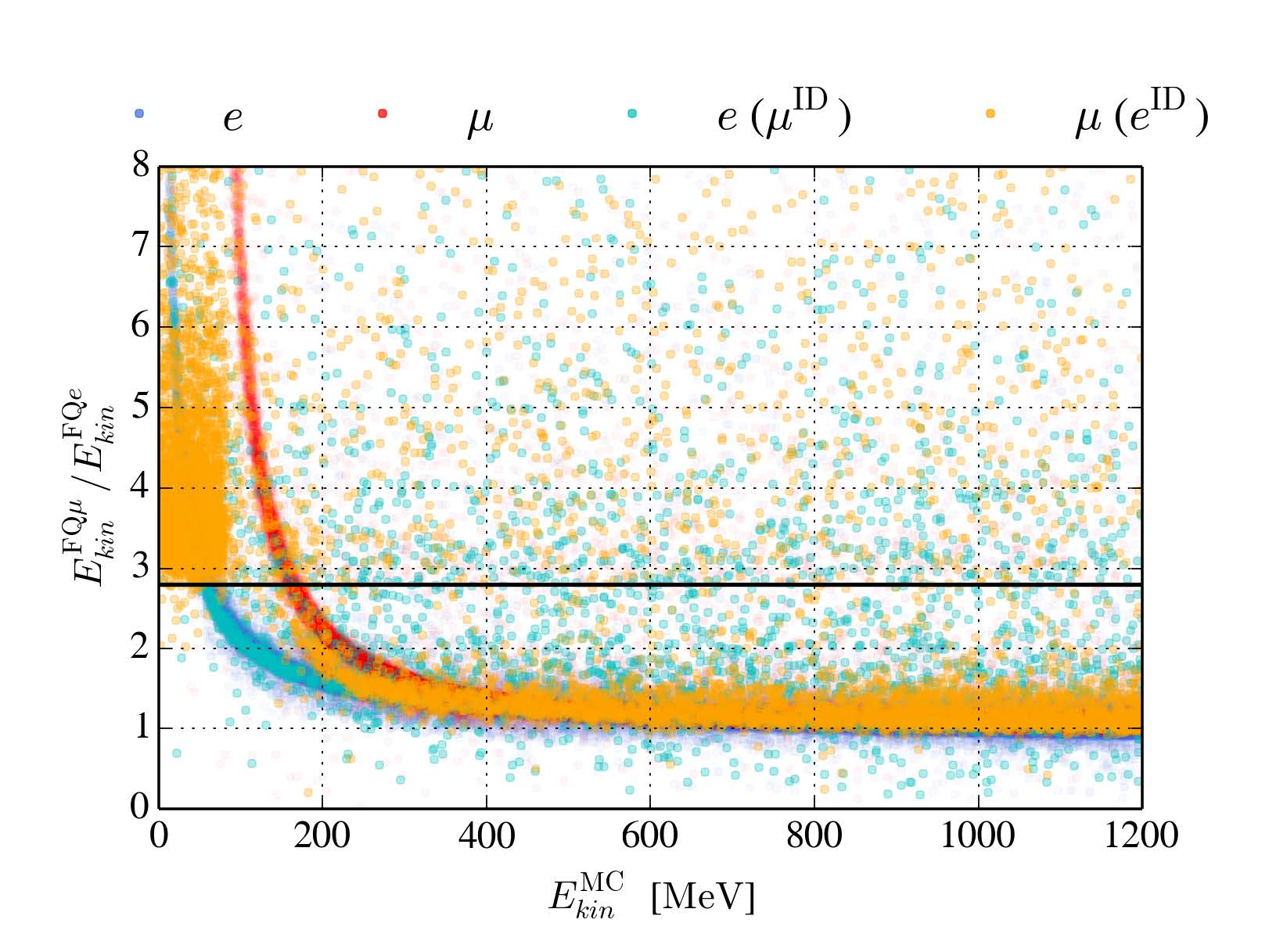}%
        \hfill
        \includegraphics[width=0.5\linewidth]{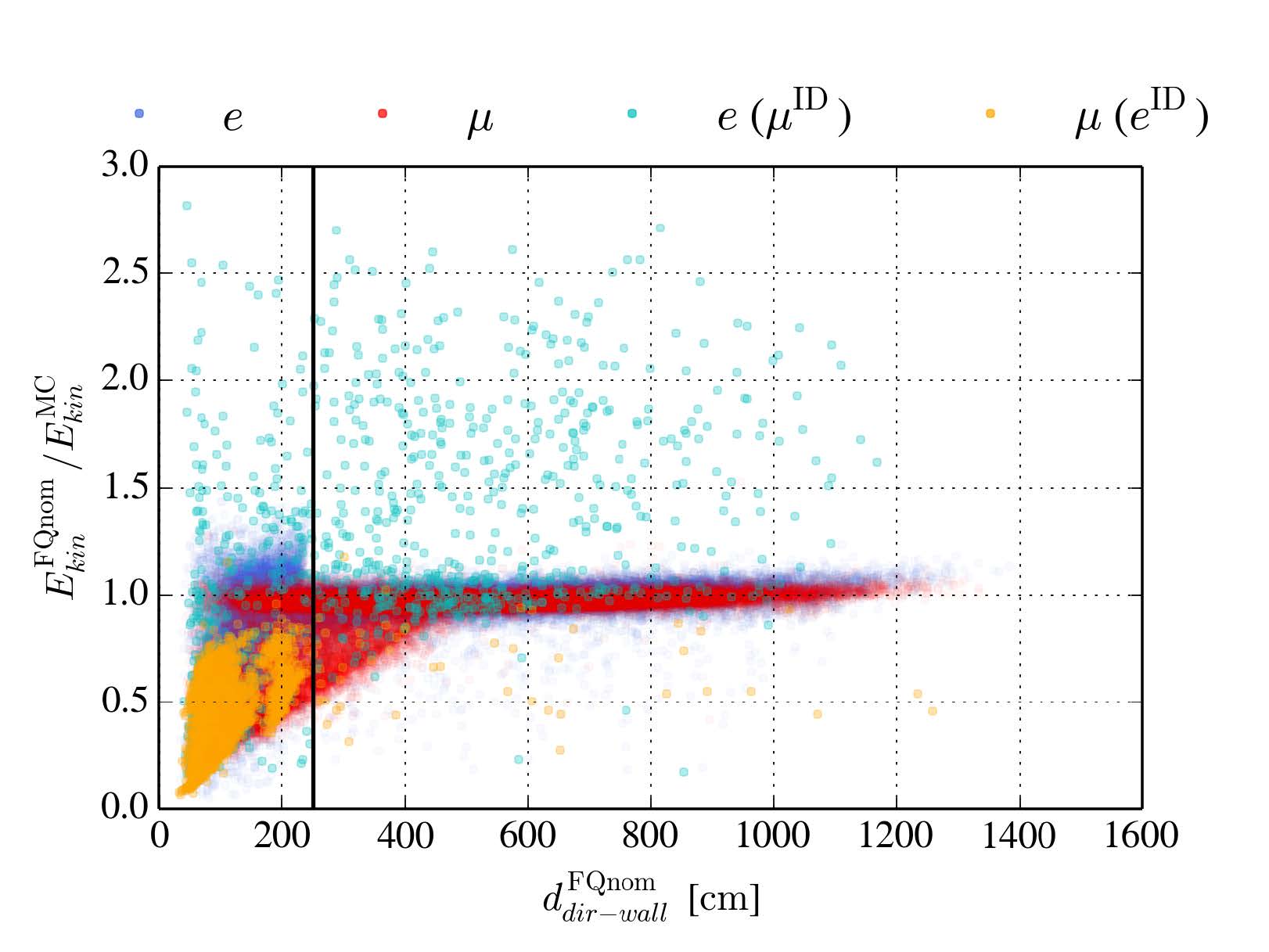}
    \end{subfigure}
    \caption{Illustration of two of the likelihood-based cuts used to remove incorrectly identified events from the charged lepton data, showing correctly identified electrons (dark blue) and muons (red) and incorrectly identified electrons ($e(\mu^{ID})$, cyan) and muons ($\mu(e^{ID})$, orange) distributed by reconstructed variables (and in one case, simulated energy). The classification uses the fiTQun-based approach described in \cite{Alekou_2022}.\\
    The sub-Cherenkov criterion (left) shows events distributed by simulated energy ($E_{kin}^\text{MC}$) and ratio between reconstructed energies using a muon ($E_{kin}^{\text{FQ}\mu}$) and electron  ($E_{kin}^{FQe}$) hypothesis, respectively. The black line indicates the cut, (removing events with $E_{kin}^{\text{FQ}\mu}/E_{kin}^{\text{FQ}e} > 2.8$), which serves to reject muon events where the muon has too little energy to produce Cherenkov radiation, and as a result is primarily detected through its electron decay product and misidentified an electron. \\
    The Cherenkov-ring resolution criterion (right) show events distributed by distance from the interaction vertex to the detector wall along the direction of the charged lepton $d^\text{FQnom}_{wall}$, and the ration the reconstructed and simulated energies. The black line indicates the cut, removing events with $d^\text{FQnom}_{wall}$ < 250 \text{cm} which are often misidentified. \\
    Figures from \cite{Alekou_2022}. } 
    \label{example_cuts}
\end{figure}

For the reconstructions in the CDR \cite{Alekou_2022}, for which the likelihood-based method was used, a number of cuts\footnote{For a full list of cuts, see \cite{Alekou_2022}, Sec. 7.2.3} are applied to the data to filter out events that are hard to reconstruct with good accuracy. These cuts are based on reconstructed variables from the fiTQun-based reconstruction and include, as example, the cuts shown in Figure \ref{example_cuts}. On the left, the ratio between reconstructed energies using a muon (neutrino) and electron (neutrino) hypothesis, respectively, is used to filter out muon (neutrino) events where the produced muons have energy below the Cherenkov threshold. These events are primarily detected via the electron from  the muon decay, and thus frequently misidentified as electron (neutrino) events. On the right, the distance from the starting initial position of the charged lepton (which would be the interaction vertex) and the detector wall along the direction of travel of the lepton, is used to filter out events where the charged lepton is produced to close to the wall, which are often misidentified.

As these cuts are based on variables from the fiTQun-based reconstruction, replacing the likelihood-based reconstruction method with a GNN-based method also means losing the ability to perform these cuts. Additionally, about half of the events are lost when applying these cuts, so removing or loosening the cuts could provide better statistics. Thus, part of this work is also investigating if the GNN classification model can perform well without these likelihood-based cuts applied. 

\section{Results}

\subsection{Charged Lepton Events}

\begin{figure}[!t]
    \centering
    \includegraphics[width=\linewidth]{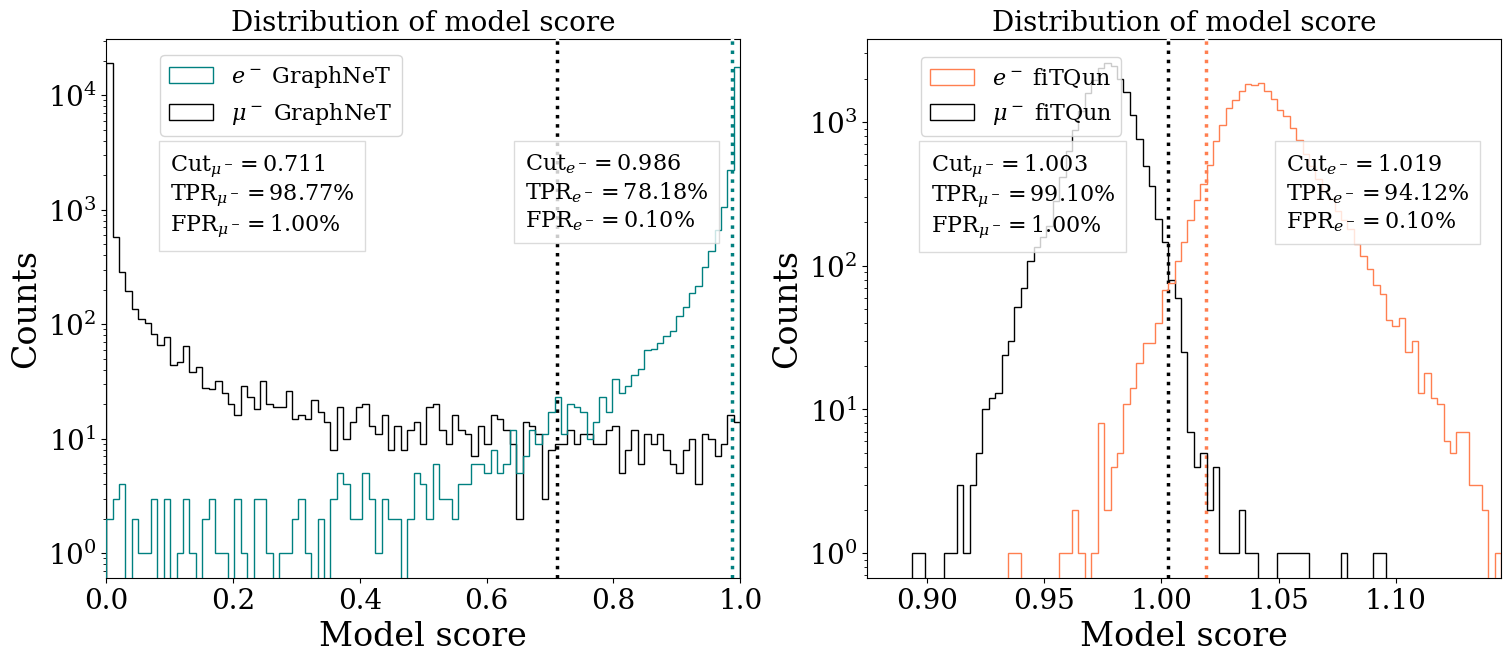}
    \caption{Model score distributions for the GNN (left) and fiTQun (right) lepton flavour classification method applied to the charged lepton events (with likelihood-based cuts). Dashed lines represent thresholds corresponding to target FPRs, and the resulting TPR is shown.}
    \label{lep_score_gnn_llh}
\end{figure}

\begin{figure}[!t]
    \centering
    \includegraphics[width=\linewidth]{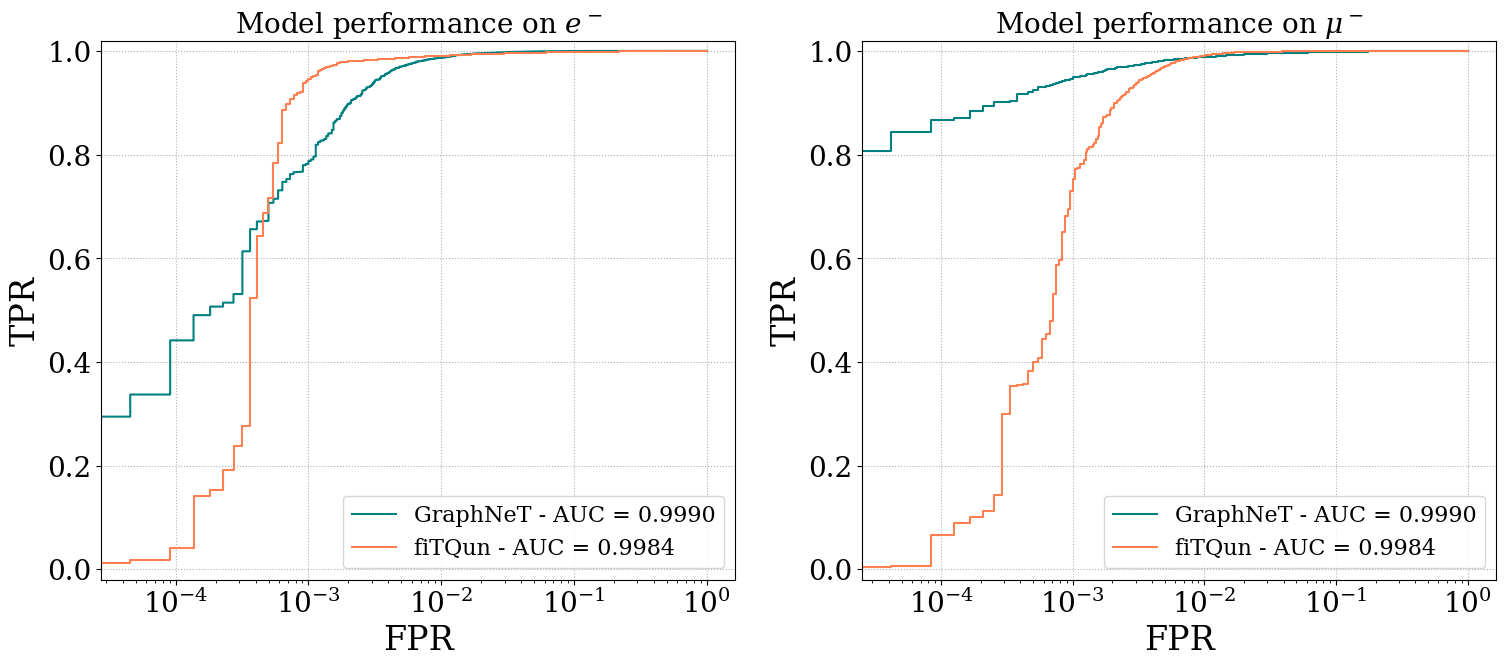}
    \caption{ROC curves for the GNN and fiTQun lepton flavour classification methods applied to the charged lepton events (with likelihood-based cuts) with electrons as signal (left) and muons as signal (right).}
    \label{lep_ROC}
\end{figure}

For charged lepton event simulations, the likelihood-based cuts described in \ref{sec:cuts} were applied to the data set to have a baseline case to compare both with the CDR results and subsequent studies in this paper. The resulting sample should have a large fraction of unambiguous events that can be easily classified by both reconstruction methods. The model score distribution of the charged lepton event classification is shown in Figure \ref{lep_score_gnn_llh} for the GNN (left) and fiTQun (right). Both the fiTQun and GraphNeT classification methods produce acceptable results on par with the performance described in the CDR, with the fiTQun-based method having slightly better performance, yielding TPRs of 99.10\% and 94.12\% for the chosen FPR rates for muons and electrons, respectively, against the 98.77\% and 78.18\% TPRs of the GraphNeT-based method. This is to be expected, since this is the type of events the likelihood-based method is optimized for. 

Figure \ref{lep_ROC} shows Receiver Operating Characteristic (ROC) curves for both reconstruction method. ROC curves are obtained by varying the model score threshold and plotting the obtained FPR (x-axis) and TPR (y-axis) and thus illustrate the trade-off between the FPR and TPR for a binary classification model. The shape of the ROC curve with the electron flavour as signal (Figure \ref{lep_ROC}, left) indicates that for looser constraints on FPR ($ > 5 \cdot 10^{-4}$) the fiTQun-based method yields greater TPRs than the GNN, while for lower on FPRs, the GNN starts to outperform the fiTQun-based method. For muons (Figure \ref{lep_ROC}, right) the performance of the two methods is similar above FPR $=10^{-2}$, while for lower target rates, the GNN has superior performance. Figure \ref{lep_ROC} also shows the calculated area under the ROC curves (AUC). The AUC describes a classification method's ability to yield a high TPR at the full FPR range, with a perfect ROC curve being a 90\textdegree angle (or a horizontal line when the FPR-axis is logarithmic) with AUC = 1. The AUC is useful as a single statistic to compare the performance of reconstruction methods, but for more rigorous testing one should focus on the specific FPR requirements.

\subsection{Charged Current Neutrino Events}

\subsubsection{With Likelihood-based Cuts}

\begin{figure}[!t]
    \centering
    \includegraphics[width=\linewidth]{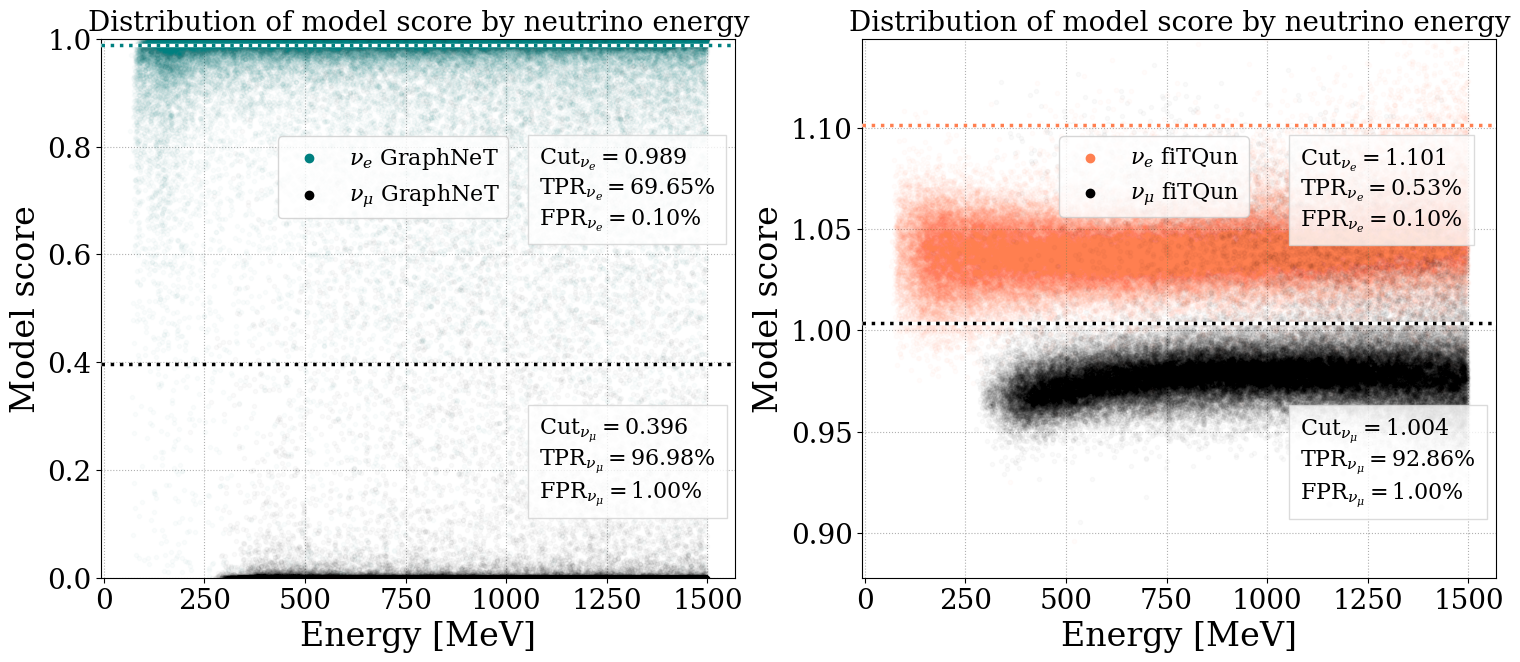}
    \caption{Model score distributed by true (simulated) neutrino energy for the GNN (left) and fiTQun (right) neutrino flavour classification methods applied to CC neutrino events with likelihood-based cuts applied. Dashed lines represent thresholds corresponding to target FPRs, and the resulting TPR is shown.}
    \label{nu_cuts_energy_gnn_llh}
\end{figure}

\begin{figure}[!t]
    \centering
    \includegraphics[width=\linewidth]{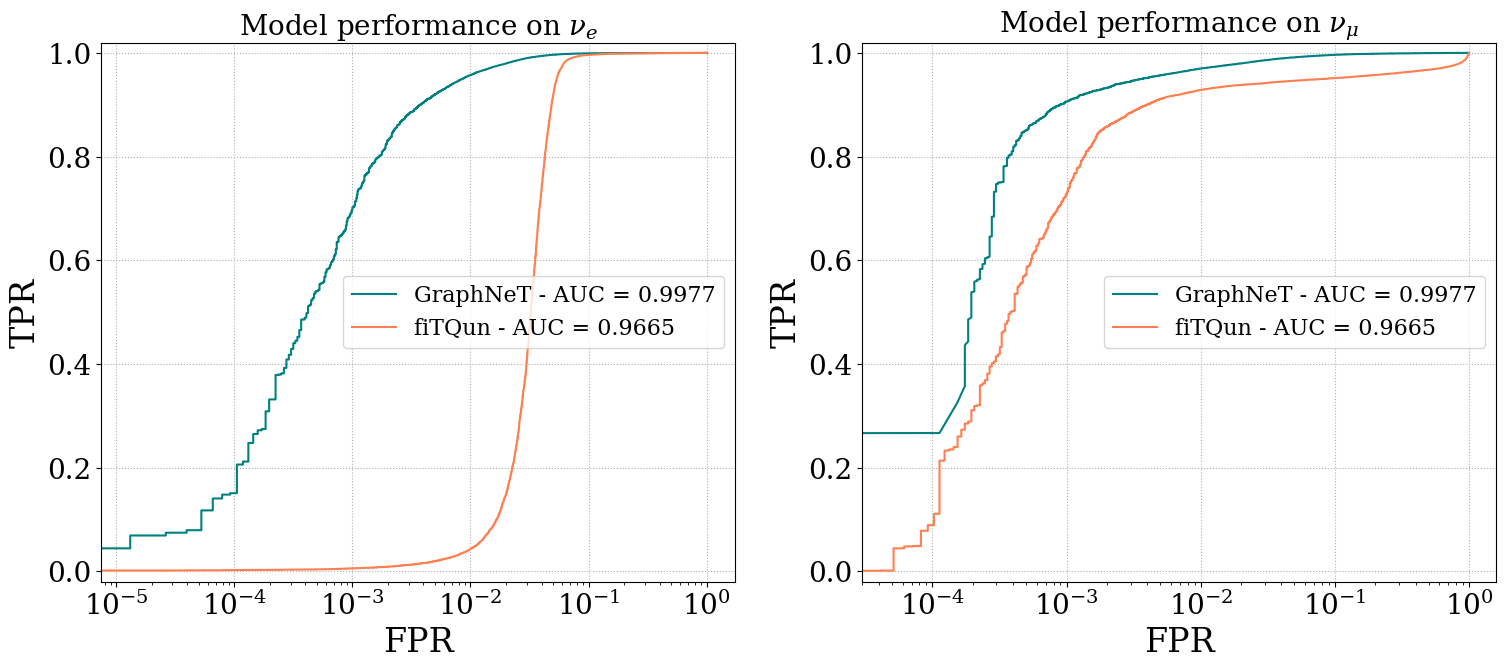}
    \caption{ROC curves for the GNN and fiTQun neutrino flavour classification methods applied to CC neutrino events with likelihood-based cuts applied with electron neutrinos as signal (left) and muon neutrinos as signal (right). Similarly to the previous data sets, the ROC curves become jagged in the low FPR region, due to low statistics, but notably the ROC curve for the GNN on $\nu_\mu$ becomes flat at around 0.01\% FPR, indicating the about 25\% of the muon neutrinos receive such a high model score that all the electron neutrinos can be filtered out without losing this fraction of muon neutrinos.}
    \label{nu_cuts_ROC}
\end{figure}

In this section the performance of the GNN and fiTQun-based classification methods applied to full neutrino interaction event simulations with the likelihood-based cuts described in \ref{sec:cuts} applied is shown, in order to illustrate how each method performs on neutrino simulations under optimal conditions. 

The model score for the neutrino event classification distributed by true (simulated) neutrino energy is shown in Figure \ref{nu_cuts_energy_gnn_llh} for the GNN (left) and fiTQun (right). The GNN-based classification yields TPRs of 96.98\% and 69.65\% for muon and electron neutrinos respectively, when applying the same restrictions on FPR, which is comparable to the performance of the reconstruction described in the CDR \cite{Alekou_2022}. The likelihood-based classification yields a similar TPR of 92.86\% for muon neutrinos, but a significantly lower TPR of 0.53\% for electron neutrinos. Although the two groups seem well separated by the likelihood-based method with no clear structure to the misidentified events, the 0.1\% FPR for electron neutrinos is too tight to produce a useful TPR.

Figure \ref{nu_cuts_ROC} shows ROC curves for both reconstruction methods, confirming that the GNN consistently outperforms the fiTQun-based method for all relevant FPR rates, and that the performance difference is the greatest for lower ($10^{-2}$ - $10^{-4}$) FPR ranges of electron neutrino events. This suggests that there is a large group of muon neutrino events that are falsely identified as electron neutrinos by fiTQun, but that the GNN can correctly identify. We return to this point in Section \ref{sec:modelcompare}.

\subsubsection{Without Likelihood-based Cuts}

\begin{figure}[!t]
    \centering
    \includegraphics[width=\linewidth]{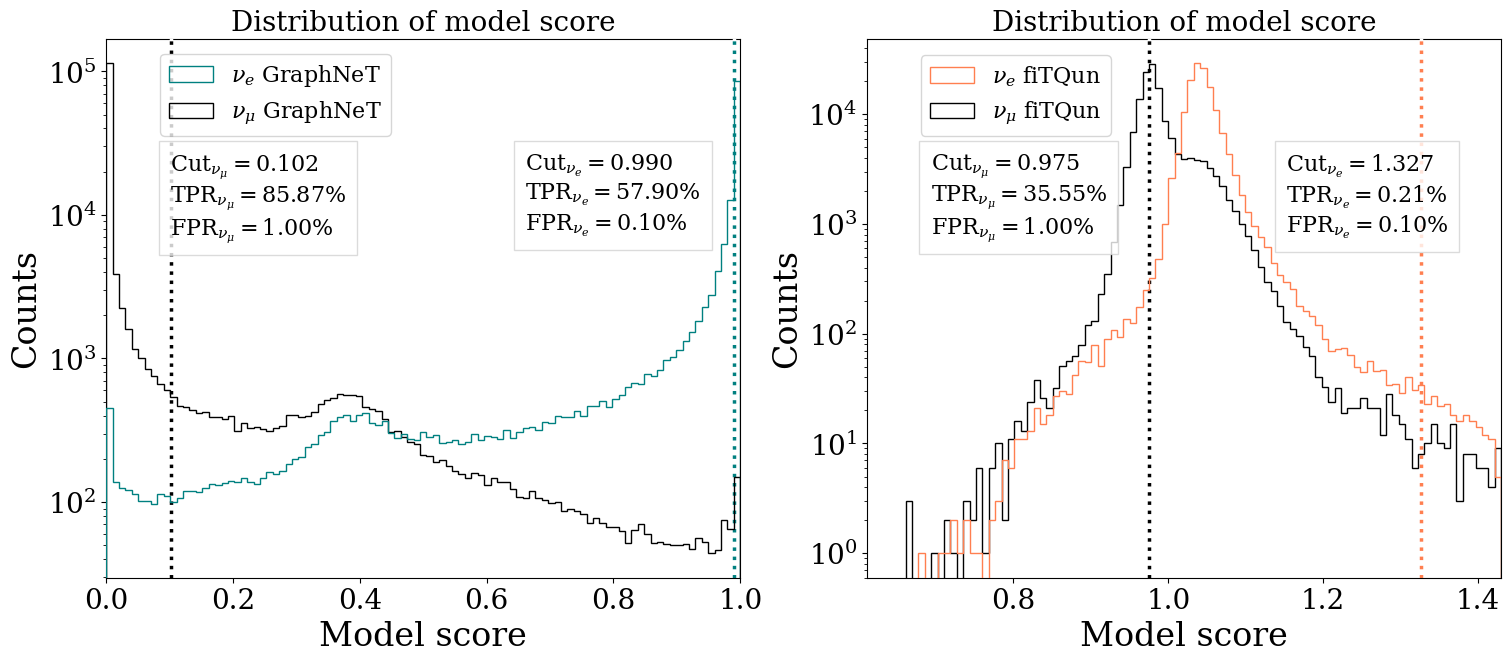}
    \caption{Model score distributions for the GNN (left) and fiTQun (right) neutrino flavour classification methods applied to CC neutrino events without likelihood-based cuts applied. Dashed lines represent thresholds corresponding to target FPRs, and the resulting TPR is shown.}
    \label{nu_score_gnn_llh}
\end{figure}

\begin{figure}[!t]
    \centering
    \includegraphics[width=\linewidth]{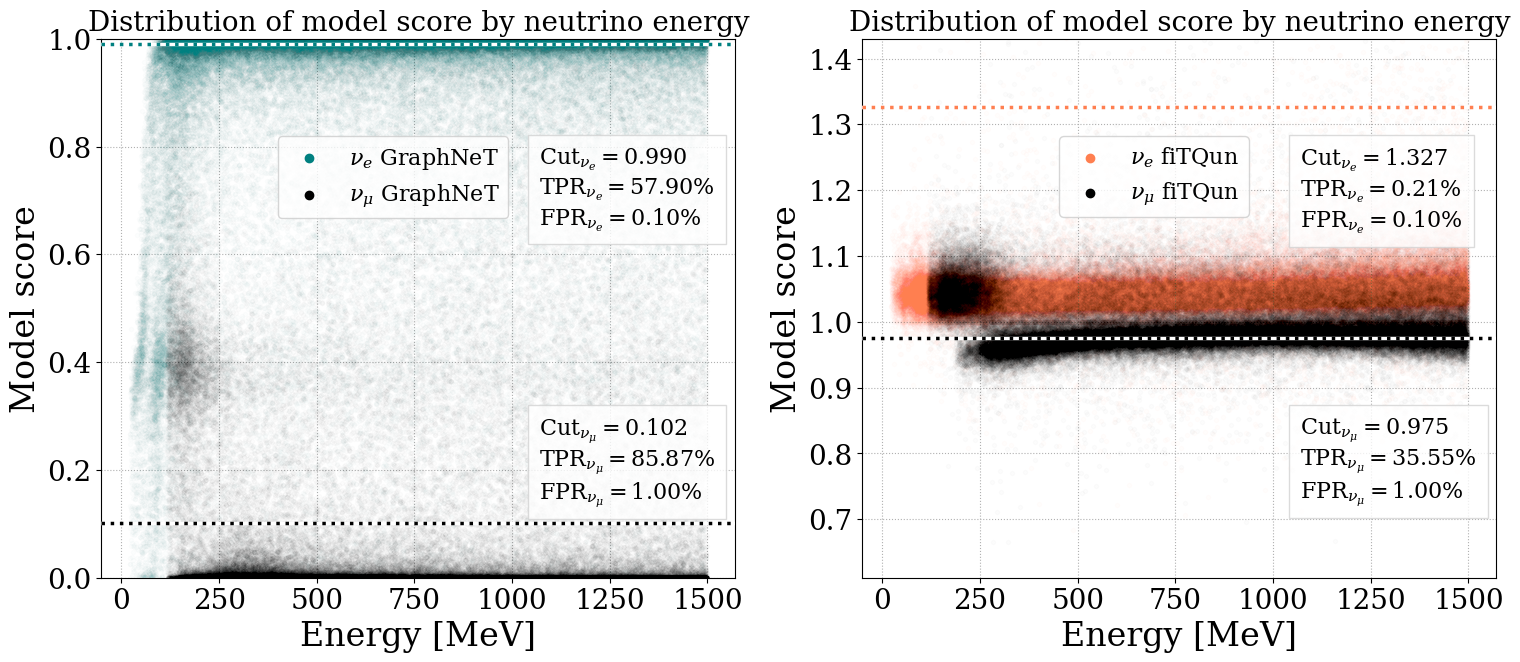}
    \caption{Model score distributed by true (simulated) neutrino energy for the GNN (left) and fiTQun (right) neutrino flavour classification methods applied to CC neutrino events without likelihood-based cuts applied. Dashed lines represent thresholds corresponding to target FPRs, and the resulting TPR is shown.}
    \label{nu_score_energy_gnn_llh}
\end{figure}

\begin{figure}[!t]
    \centering
    \includegraphics[width=\linewidth]{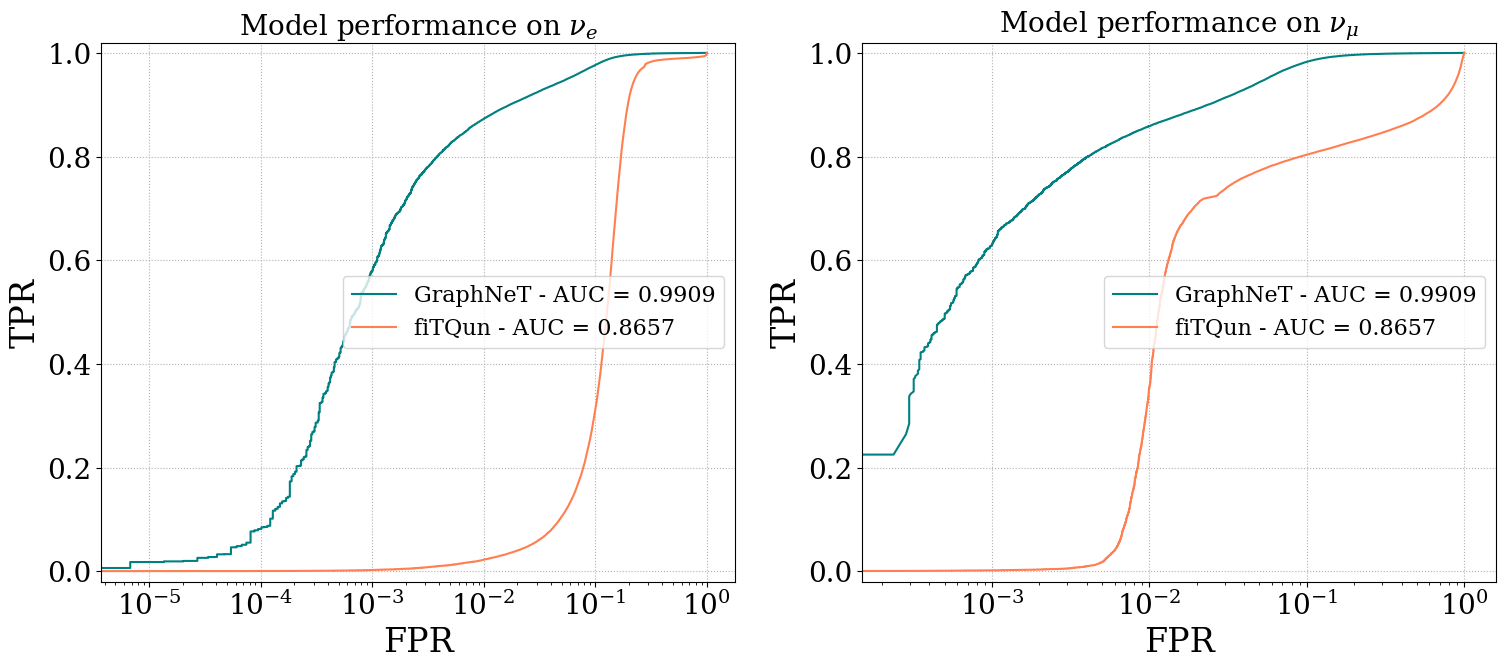}
    \caption{ROC curves for the GNN and fiTQun neutrino flavour classification methods applied to CC neutrino events without likelihood-based cuts applied with electron neutrinos as signal (left) and muon neutrinos as signal (right).}
    \label{nu_ROC}
\end{figure}

In the following, the likelihood-based cuts described in \ref{sec:cuts} were disregarded to investigate if the GNNs can be applied to the data without these cuts, which as argued before is important if the GNNs are to be used to optimise the detector geometry. The model score distributions of the neutrino event classification is shown in Figure \ref{nu_score_gnn_llh} for the GNN (left) and fiTQun (right). For this sample, the performance of both methods is similar to the performance with the likelihood-based cuts applied, though it is worse in all cases. The GNN classification yields TPRs of 85.87\% and 57.90\% for muon and electron neutrinos respectively, and the likelihood-based classification yields lower TPRs of 35.55\% and 0.21\% for muon and electron neutrinos respectively.

The model score distributed by true (simulated) neutrino energy shown in Figures \ref{nu_score_energy_gnn_llh} for the GNN (left) and fiTQun (right), shows a clear band of muon neutrinos in the lowest energy range, that overlaps completely with the electron neutrinos, making it hard to exclude them without losing the majority of electron neutrinos. Since muons produced in interactions in this range will have low enough energy that the dominant part of their event signature will be from the electrons produced from their decay, such a band is expected. While the GNN performance also appears to be worst in the low energy range, such a band is not present. Instead, many of the ambiguous events obtain a model score of about 0.4, leading to them being excluded when applying the restrictions on FPR. This indicates that the GNN, from being exposed to enough of these events during training, can learn to identify them as ambiguous and assign them a model score that represents this.

Figure \ref{nu_ROC} shows ROC curves for both methods, confirming again that the GNN consistently outperforms the fiTQun-based method, and that while the difference in performance is still greatest for lower FPR electron neutrino events, the performance difference for muon neutrino events is greater than with the likelihood-based cuts applied.

\subsubsection{Performance Comparison}\label{sec:modelcompare}

\begin{figure*}[!t]
    \centering
    \includegraphics[width=\linewidth]{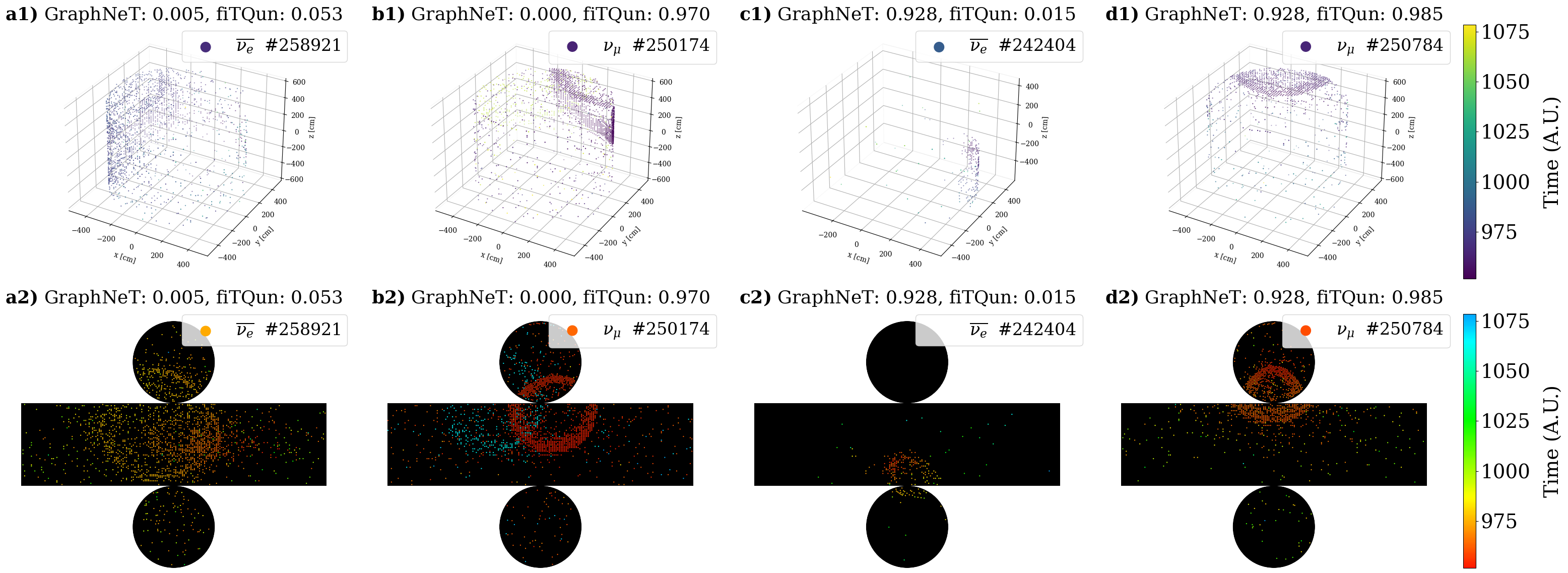}
    \caption{3D (top) and 2D (bottom) event displays of events with different combinations of Confidence-Correctness Scores for each neutrino flavour classification method as explained in the text. The event displays show PMT signals distributed by the spatial (x, y, z) position of the PMT, with the colour indicating the arrival time of the signal.}
    \label{discrp_event_disp_fTime}
\end{figure*}

Many factors impact the performance of the the reconstruction methods applied to the complex neutrino interaction events. In the following, a few examples of events where the two methods have similar or dissimilar performance are inspected. While the examples do not exhaust all the relevant event topologies, they can provide valuable insight about features to consider when designing the ESS$\nu$SB reconstruction pipeline. Figure \ref{discrp_event_disp_fTime} shows event displays in 3D (top) and 2D (bottom) of selected neutrino events, created to investigate the difference in performance on neutrino data between the two methods. The events were selected by the difference between the model score and the truth label (0 for $\nu_\mu$, 1 for $\nu_e$), here denoted the Confidence-Correctness Score (CCS), for each method. The CCS, illustrated in Eq. \ref{CCScalc}, is a measure of how "mistaken" the classification method is for a given event, regardless of the flavour of the event. A CCS of $\approx0$ (a correctly confident classification) can either be a $\nu_\mu$ event (0) with a model score of $\approx0$, or a $\nu_e$ event (1) with a model score of $\approx1$. In contrast, a CCS of $\approx1$ signifies a confidently incorrect prediction, and arises from a model score of $\approx0$ for a a $\nu_e$ event (1) or vice versa. In the following, the events where selected to have different combinations of CCS's for the two reconstruction methods\footnote{Combinations of: low-low, low-high, high-low, high-high, where a low score ($CCS<0.1$) indicates a correct prediction with high confidence, and a high score ($CCS>0.9$) indicates an incorrect prediction with high confidence. The value 0.1 was chosen to accept most of the unambiguous events and the value 0.9 = 1-0.1 is derived from that.}.

\begin{align}
    CCS = \mid \text{truth} - \text{prediction}\mid
    \label{CCScalc}
\end{align}

The first event from the left (Figure \ref{discrp_event_disp_fTime}, a1 and a2) shows an event where both methods predict correctly with great confidence. This event has many detector hits, providing lots of information to the reconstruction algorithms, as well as a clear ring structure. The second event (Figure \ref{discrp_event_disp_fTime}, b1 and b2), where the GNN-based method predicts correctly and the fiTQun-based method incorrectly, is of particular interest, as both the 2D and 3D event displays show two distinct Cherenkov Rings with a significant time delay, most likely from a event where a muon neutrino produces a muon that decays and produces an electron. This indicates that the GNN has learned to identify these more complex events out of the box, even for cases where it is difficult for fiTQun, a finding that is consistent when looking at Figure \ref{gb_event_disp_fTime} of Appendix \ref{sec:AEventDisplays}, which shows several event displays of similar CCS combination, many of which exhibit double rings. The third event (Figure \ref{discrp_event_disp_fTime}, c1 and c2) where the fiTQun-based method predicts correctly with great confidence, and the GNN predicts incorrectly, has few detector hits, but a clear ring structure. This can be interpreted as the event not having enough data points for the GNN - which as an ML model needs more information to function optimally - to identify the pattern, but ring-like enough for the fiTQun-based method - which as an LLH-based method does not necessarily have the same limitations - to converge. The final event (Figure \ref{discrp_event_disp_fTime}, d1 and d2) has an average amount of detector hits, and a clear ring shape, and should be easily interpretable by both methods. It is not clear why both methods have poor performance on such an event, and future studies should look into this behaviour. Since the main focus of this section is on the events where the methods differ in performance, it is not investigated further here.

\subsubsection{Pion Production}\label{sec:pions}

\begin{figure}[!t]
    \centering
    \includegraphics[width=\linewidth]{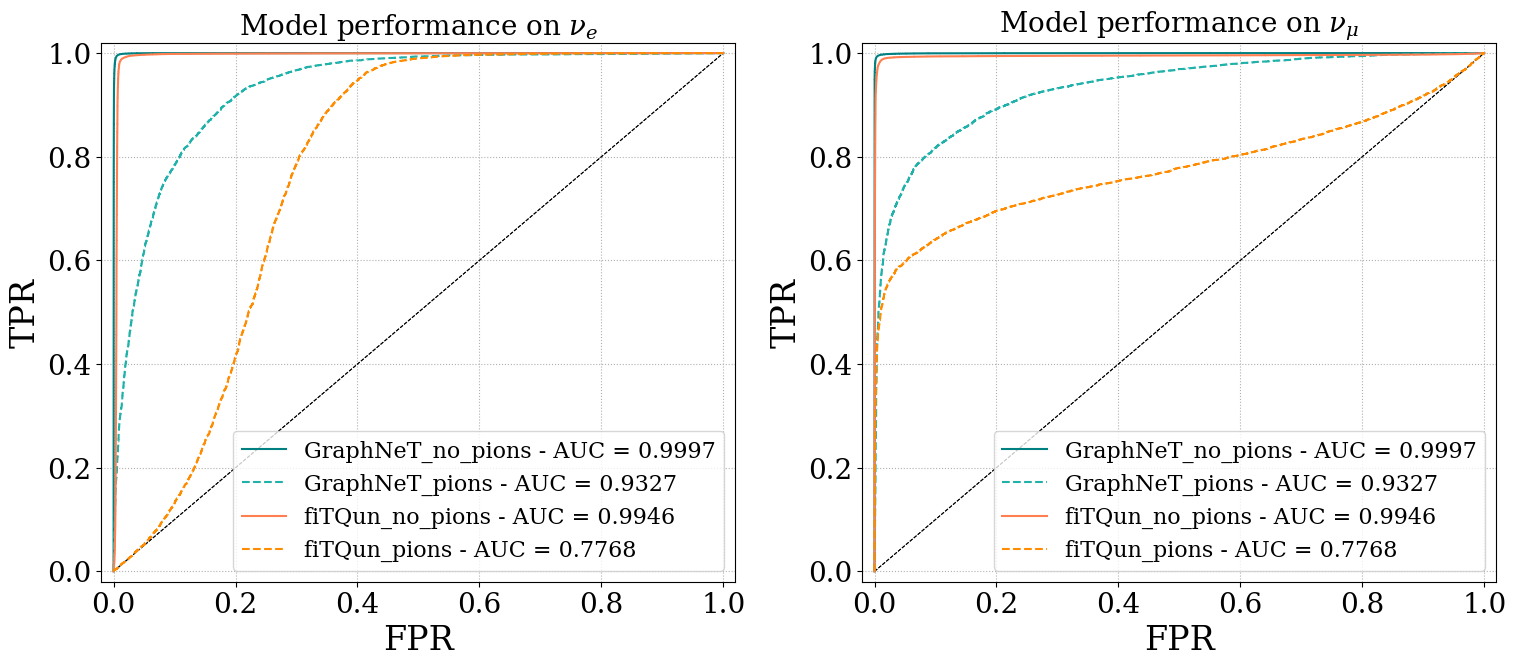}
    \caption{ROC curves for the GNN and fiTQun neutrino flavour classification methods applied to CC neutrino events (with likelihood-based cuts) with electron neutrinos as signal (left) and muon neutrinos as signal (right) in the cases where pions and no pions are produced in the final state.}
    \label{pions_ROC}
\end{figure}

\begin{figure}[!t]
    \centering
    \includegraphics[width=\linewidth]{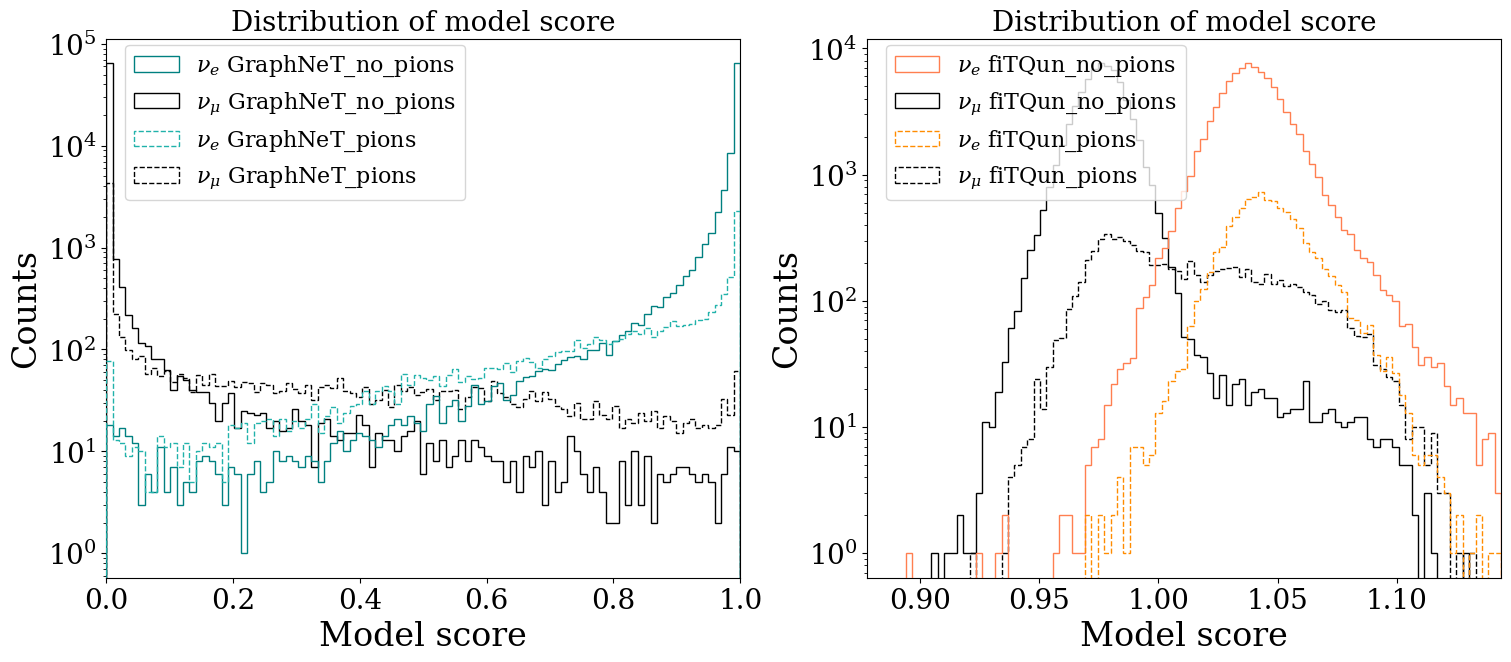}
    \caption{Model score distributions for the GNN (left) and fiTQun (right) neutrino flavour classification methods applied to CC neutrino events (with likelihood-based cuts) in the cases where pions and no pions are produced in the final state.}
    \label{pion_scorehist_log}
\end{figure}

Another factor that blurs the picture of neutrino interactions in the ESS$\nu$SB detector is the production of pions. Pions are a common product of weak neutrino-nucleus interactions, created both in resonance production processes and coherent scattering. Both CC and NC coherent scattering can produce pions, with CC processing being the dominant contributor. The pion production can interfere with measurements, both because they can create a signature that looks like Cherenkov radiation from a charged lepton, and because their energy will be missing from the charged lepton energy reconstruction and consequently from the neutrino energy estimation \cite{ParticleDataGroup:2024cfk, REIN198179}. 

Figure \ref{pions_ROC} shows the ROC curves for the GNN and fiTQun-based methods applied to the neutrino events (in this case with the likelihood-based cuts, to better visualise the differences in performance between the two methods), but separated into events where pions are produced in the final state (which make up $\sim$10\% of the events) and events where no pions are produced. It is clear, that both methods are able to classify events with no pion production, whereas for the event with pion production, the fiTQun-based method performs significantly worse than the GNN for both electron and muon neutrinos. This is also demonstrated in Figure \ref{pion_scorehist_log} which show histograms of the model scores for the GNN (left) and fiTQun (right) classification methods. For the GNN, there is still a good separation of the events with pions, but the number of events receiving a score near 0 or near 1 is lower, while the number of events receiving scores in the rest of the spectrum is higher, signifying that the events are harder to classify. For the fiTQun-based method, the distribution of $\nu_e$ events is unchanged, but the distribution of $\nu_\mu$ events are shifted to the right (higher model scores), resulting in a large overlap of $\nu_e$ and $\nu_\mu$ events. This indicates, that one might want to start by using classification model (like another GNN) to separate events into classes with or without pions before running the reconstruction, in order to treat the two cases separately. We will test this idea in the next subsection.

\subsubsection{Pion Classifier}\label{sec:pionclass}

\begin{figure}[!t]
    \centering
    \includegraphics[width=\linewidth]{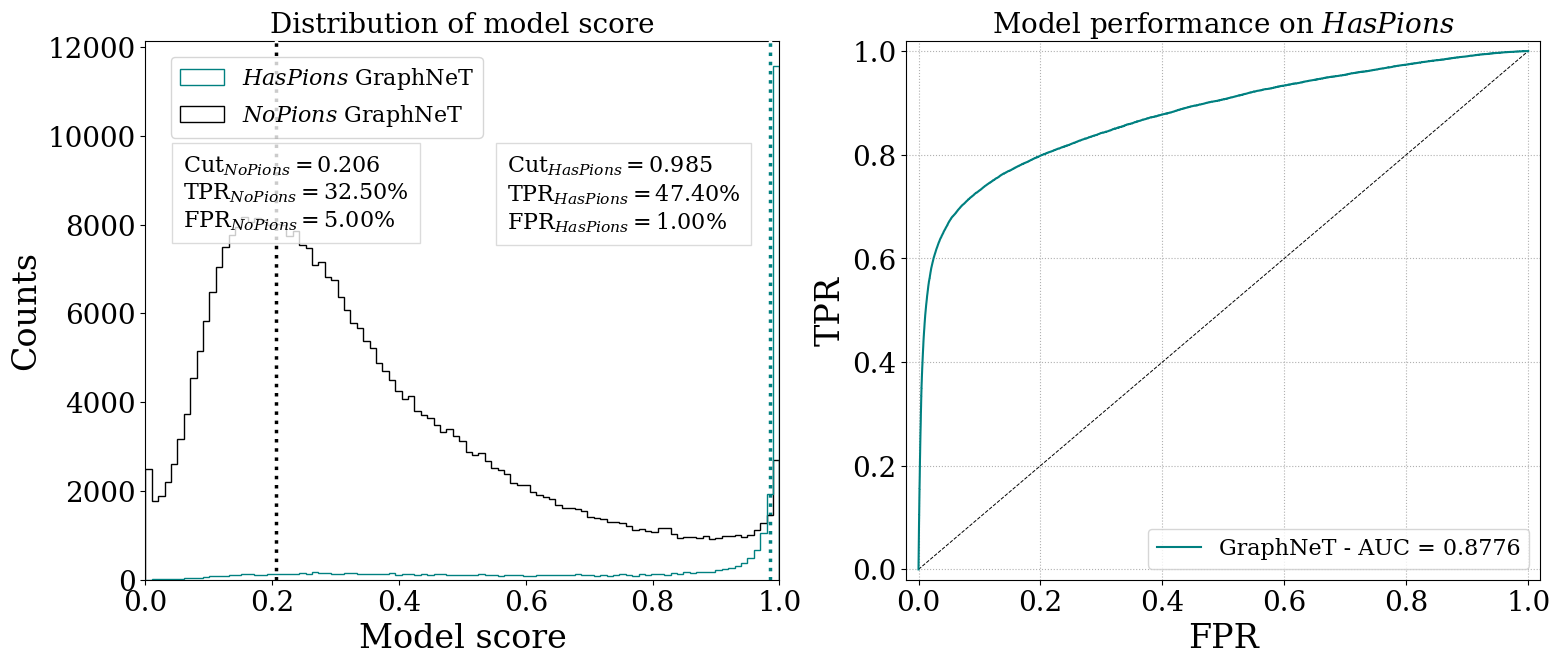}
    \caption{Model score distribution (left) and ROC curve (right) for the GNN pion production classification methods applied to CC neutrino events. Dashed lines on the left represent cuts corresponding to chosen FPRs, and the resulting TPR is shown.}
    \label{pionmodel_scorehist_ROC}
\end{figure}

Following the results in Section \ref{sec:pions}, a GNN classification model was trained on the CC (CC) neutrino interaction data set to distinguish between events with and without pion production. The model target variable was chosen to be a boolean that had the value true (1) for events where any number of pions were created in the interaction, and false (0) for events where no pions where produced. Neutrino flavour was not taken into account for this model, and the likelihood-based cuts were not applied, as the goal was to obtain the most general model possible to separate all types of events by pion production. The model score distribution and the ROC curve for the resulting model, are shown in Figure \ref{pionmodel_scorehist_ROC} on the left and right side, respectively. Together, the figures show that while the model can classify most events correctly, it is harder for the model to confidently identify events without pion production than it is for events with pion production. This can be seen from the model score distribution on the left, as almost all the events with pion production are assigned a model score close to 1, whereas the events without pion production receive a wide range of model scores. Consequently, for a FPR requirement of 1.00\%, the model can retain 47.4\% of the sample of events with pions. For events without pion production, the majority of the sample is lost when enforcing an FPR of 1.00\%. Instead, Figure \ref{pionmodel_scorehist_ROC} shows the result of raising the requirement to 5.00\% FPR, which produces allow a TPR of 32.5\%. The impact of using this classifier will be investigated in the following section.

\subsubsection{Flavour Classification Performance Split by the Pion Model}

\begin{figure}[!t]
    \centering
    \includegraphics[width=\linewidth]{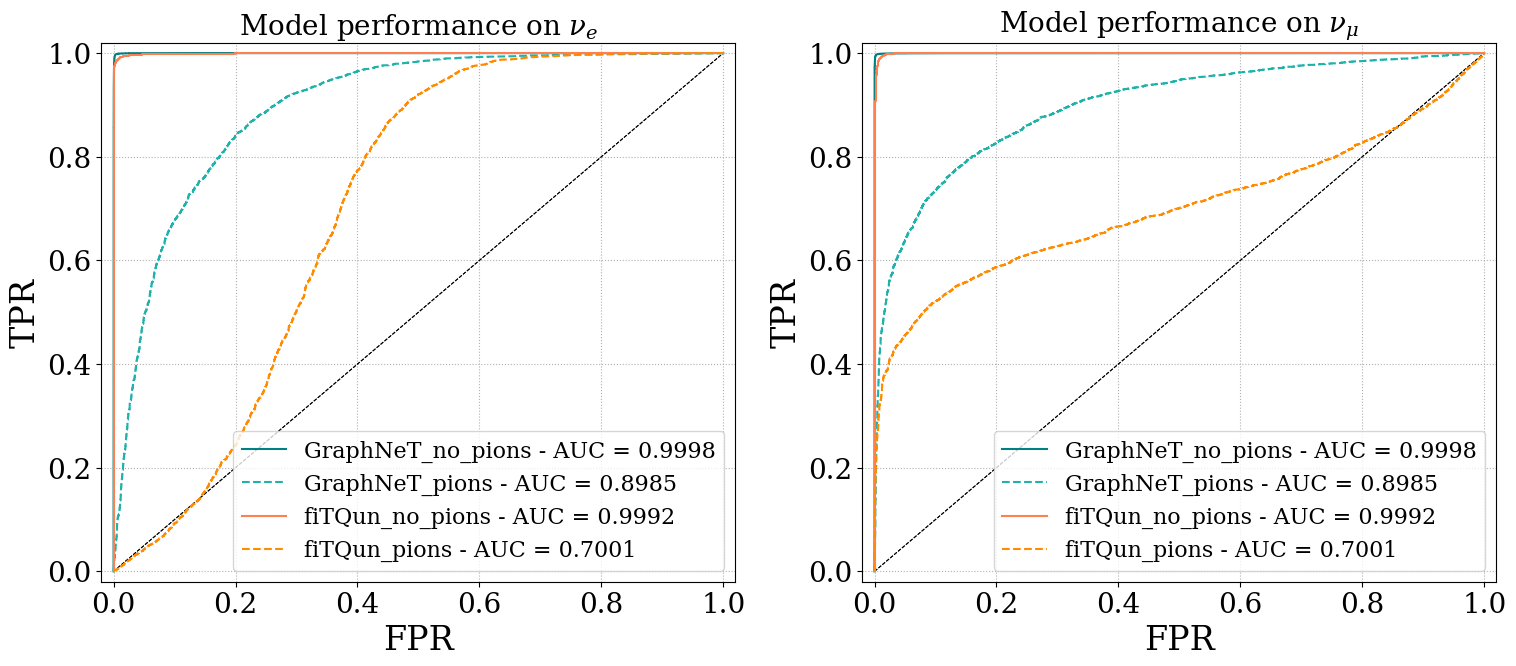}
    \caption{ROC curves for the GNN and fiTQun neutrino flavour classification methods applied to CC neutrino events (with likelihood-based cuts) with electron neutrinos as signal (left) and muon neutrinos as signal (right) separated by events with and without pion production using the pion production classification model.}
    \label{modelpions_ROC}
\end{figure}

\begin{figure}[!t]
    \centering
    \includegraphics[width=\linewidth]{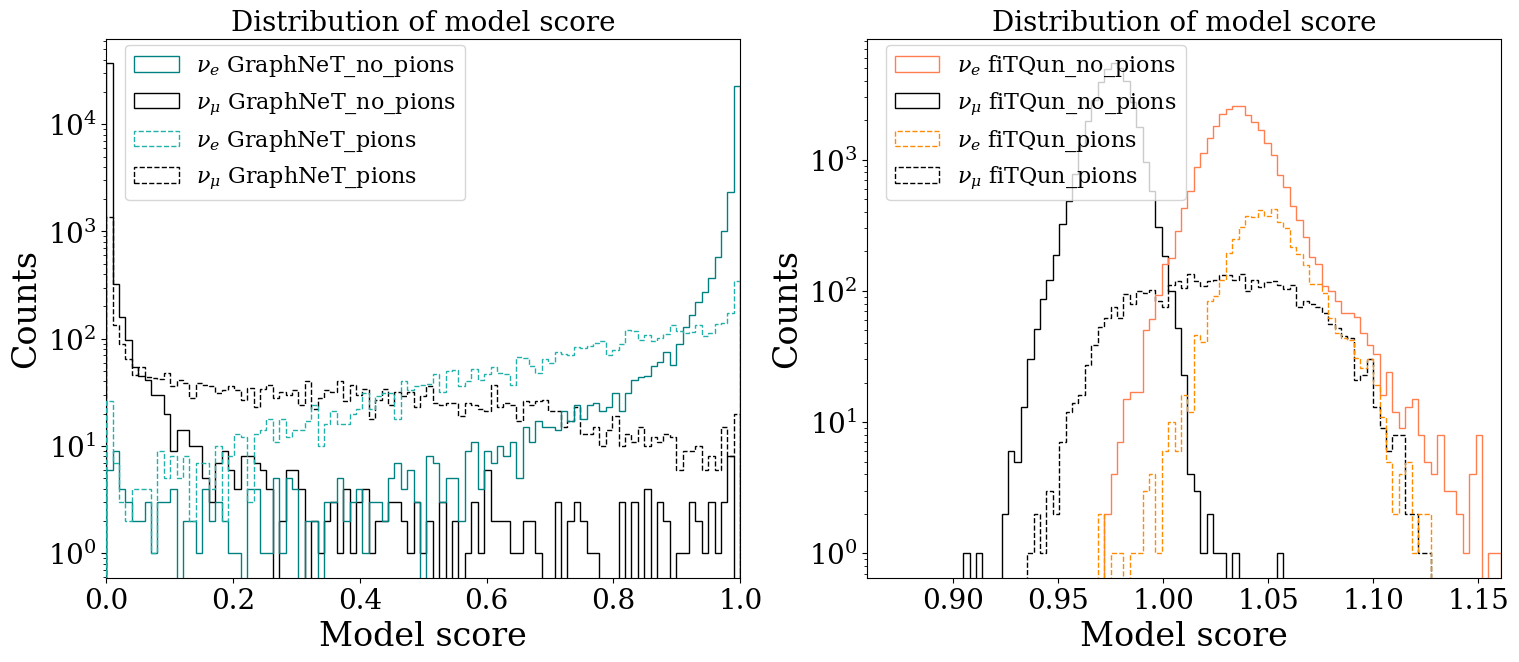}
    \caption{Model score distributions for the  for the GNN (left) and fiTQun (right) neutrino flavour classification method applied to CC neutrino events (with likelihood-based cuts) separated by events with and without pion production using the pion production classification model.}
    \label{modelpion_scorehist_graphnet_log}
\end{figure}

With the pion classifier described in the previous section completed, the next objective was to study how the reconstruction for both the GNN and the fiTQun-based method was impacted by filtering the data with this classifier. For this, both the GNN and fiTQun-based flavour classification methods were applied to the neutrino events (in this case with the likelihood-based cuts, to better visualise the differences in performance between the two methods), but separated by the pion production classification model described in Section \ref{sec:pionclass}, using the thresholds shown on the left of Figure \ref{pionmodel_scorehist_ROC}, to obtain reasonably pure samples of events with and without pion production. Figure \ref{modelpions_ROC} shows the ROC curves for two models and the two data samples, which are almost identical to the ones shown in Figure \ref{pions_ROC}, showing that both methods perform very well on the events without pion production, and that the GNN has significantly better performance on the events with pion production than the fiTQun-based method. This is confirmed when looking at Figures \ref{modelpion_scorehist_graphnet_log} which shows the model score distributions for the GNN (left) and the fiTQun-based method (right). For the GNN, while  the overlap is greater for the events with pion production than for the events without, there is still a significant separation for both samples. For the fiTQun-based method, the separation of events without pions is very clear, whereas there is a large overlap between the two flavours for the events with pion production. This indicates that the pion production classifier is working as intended, separating events into categories with different reconstruction difficulty.

\subsection{Neutral Current Neutrino Events}

\begin{figure}[!t]
    \centering
    \includegraphics[width=\linewidth]{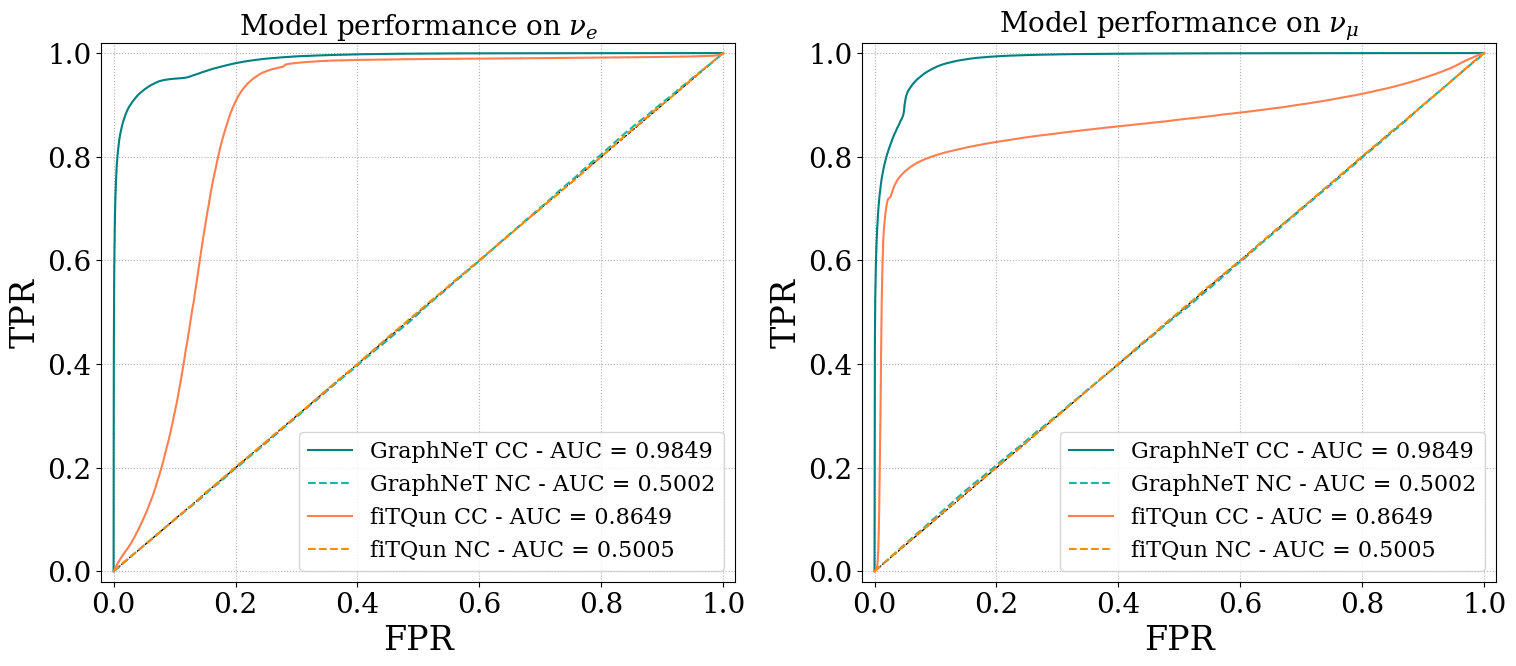}
    \caption{ROC curves for the GNN and fiTQun neutrino flavour classification methods applied to CC and NC neutrino events with electron neutrinos as signal (left) and muon neutrinos as signal (right) separated by interaction type.}
    \label{withnc_ROC}
\end{figure}

\begin{figure}[!t]
    \centering
    \includegraphics[width=\linewidth]{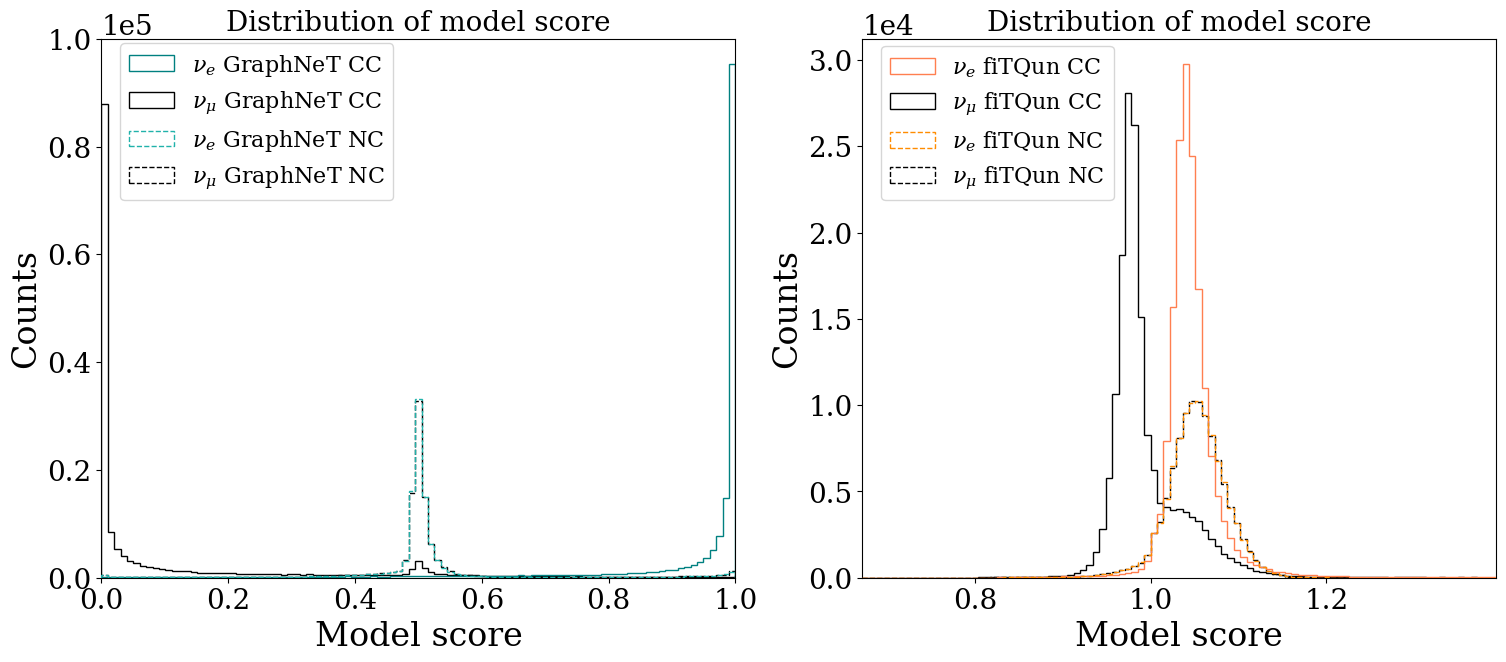}
    \caption{Model score distribution for the GNN (left) and fiTQun (right) neutrino neutrino flavour classification method applied to CC and NC neutrino events separated by interaction type.}
    \label{withnc_scorehist_nonlog}
\end{figure}

For the next section, the neutrino interaction dataset with both CC and NC events is used without any likelihood-based cuts. This includes 400.000 simulated NC events in addition to the CC events introduced in Table \ref{tab:samplesizes} and used in the previous sections. A GNN neutrino flavour classification model was trained on 100.000 events with equal representation of interaction types, flavours and neutrinos/antineutrinos, and tested on the remaining events. Figure \ref{withnc_ROC} shows the ROC curves for this GNN and the fiTQun-based method. The data for the ROC curves is split by CC and NC interactions, and Figure \ref{withnc_ROC} shows clearly, that for the NC dataset both methods have poor performance comparable to a random selection. This is also demonstrated in Figures \ref{withnc_scorehist_nonlog} which shows histograms of the model scores for the GNN (left) and fiTQun (right) classification methods, with the same data separation. For both methods there is a complete overlap in model scores for the NC events. An interesting observation is that the GNN scores almost all NC events close to 0.5, indicating that the model is unable to learn patterns that separate the events. This signifies that the development of a method for filtering out the NC events and only analyse the CC events could be beneficial.

\subsubsection{Separation of CC and NC events}

\begin{figure}[!t]
    \centering
    \includegraphics[width=\linewidth]{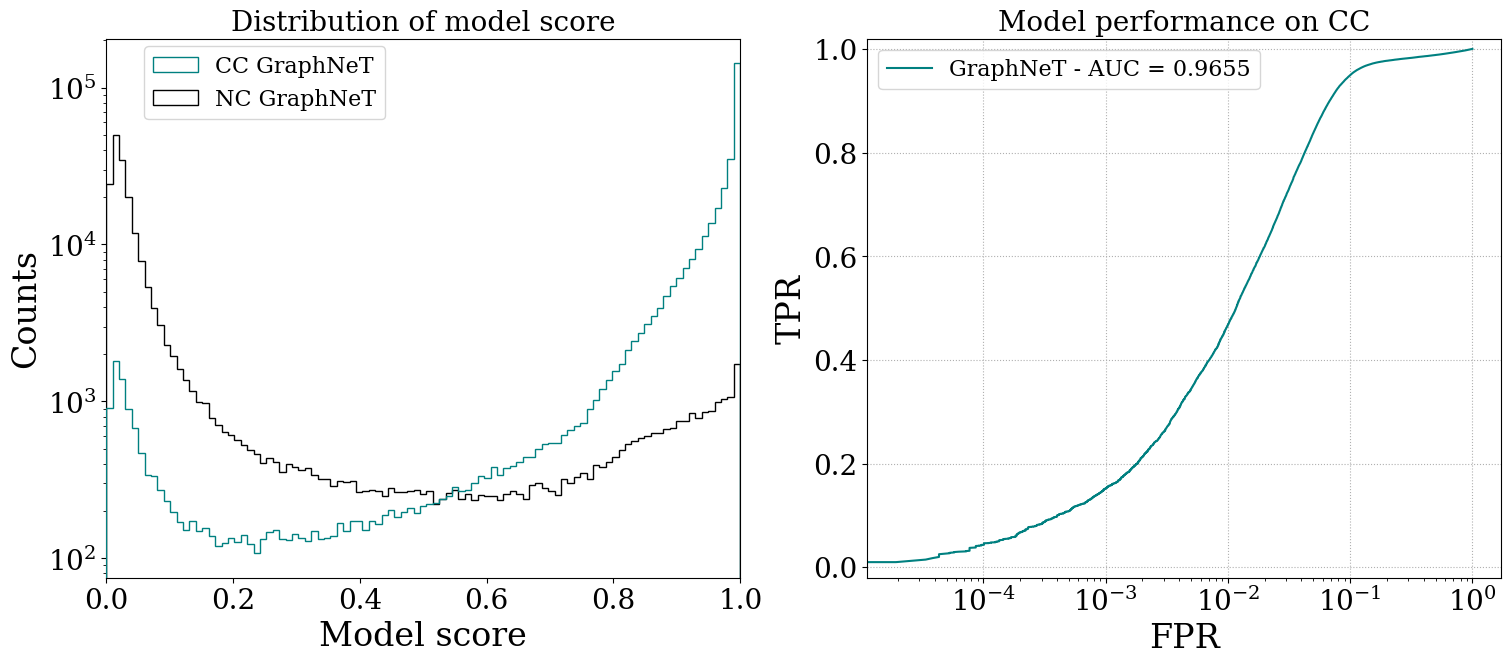}
    \caption{Model score distribution (left) and ROC curve (right) for the GNN interaction type classification method applied to CC and NC neutrino events.}
    \label{ccnc_scorehist_ROC}
\end{figure}

In an effort to meet the needs described earlier in this section, a GNN model for classifying events according to the interaction type was trained on the neutrino interaction data set with both interaction types represented, and the results are shown in Figure \ref{ccnc_scorehist_ROC}. As this is not a feature of the fiTQun-based method, no comparison is shown. The ROC curve in the right plot of Figure \ref{ccnc_scorehist_ROC} shows promising performance, as 99\% of NC events can be discarded while keeping a little less than half of the CC events. The histogram on the left of Figure \ref{ccnc_scorehist_ROC} shows a greater spread of the model score for the NC events, as well as a portion of CC events being wrongly identified. But if the only goal is a pure sample of CC events, this is irrelevant.

\subsection{Reconstruction Speed}

Recorded training and reconstruction times for the GNN are shown in Table \ref{tab:speed} and compared to the average event reconstruction time of the fiTQun framework. The benchmarking was done using the full neutrino data set of 800,000 events.

Using the GraphNeT framework involves some overhead when converting data to one of the database formats, for which the framework is optimised. But the majority of the time consumption lies in the training step. Including model training, the total reconstruction is faster than the fiTQun reconstruction by a factor of $10^3$. This trend would only improve with more events, since one only needs to train on a subset of the events, the size of which can be kept constant. 

In a scenario where one could use an already trained model to reconstruct events (for instance if one makes new simulations with only minor changes to the detector layout), the reconstruction speed improvement can reach a factor of $10^4$.

\begin{table}[!b]
    \centering
    \begin{tabular}{|ll|}
        \hline
        Data extraction & $10^{-4}$ mins/event\\
        Training & $10^{-3}$ mins/event \\
        Reconstruction & $10^{-4}$ mins/event \\
        \hline
         & \\
        fiTQun Reconstruction & 1 min/event \\
        \textbf{Improvement} & \boldmath{$10^3$} (w/ training) / \boldmath{$10^4$} (w/o training) \\
        \hline
    \end{tabular}
    \caption{Comparison of reconstruction speeds of the fiTQun and GNN classification methods, with and without training of the GNN.}
    \label{tab:speed}
\end{table}

\section{Conclusion}\label{sec13}

In this work we propose a new classification method for the ESS$\nu$SB based on Graph Neural Networks (GNNs) and the GraphNeT framework \cite{Søgaard2023}. We demonstrate that the GNN-based method has comparable performance to the fiTQun method on simulated charged lepton events with likelihood-based cuts applied to reject difficult events, and that the GNN performance surpassed that of fiTQun for simulated CC neutrino events both with and without the aforementioned cuts. 

We also present results that indicate that the GNN-based method is able to accurately reconstruct more events with two Cherenkov rings (from muons decaying to electrons), and demonstrate how GNNs can be trained to separate CC and NC events as well as events with and without pion production, and how this is beneficial, as the event types differ in reconstruction performance.

The reconstruction carried out using the GNN method is demonstrated to have a factor of $10^4$ increase in performance speed from the fiTQun method. This will be very beneficial to the ESS$\nu$SB experiment, providing great flexibility in the detector design phase.


\appendix
\section{Events Displays of Events with Superior GNN Performance}\label{sec:AEventDisplays}

\begin{figure*}[!b]
    \centering
    \includegraphics[width=.91\linewidth]{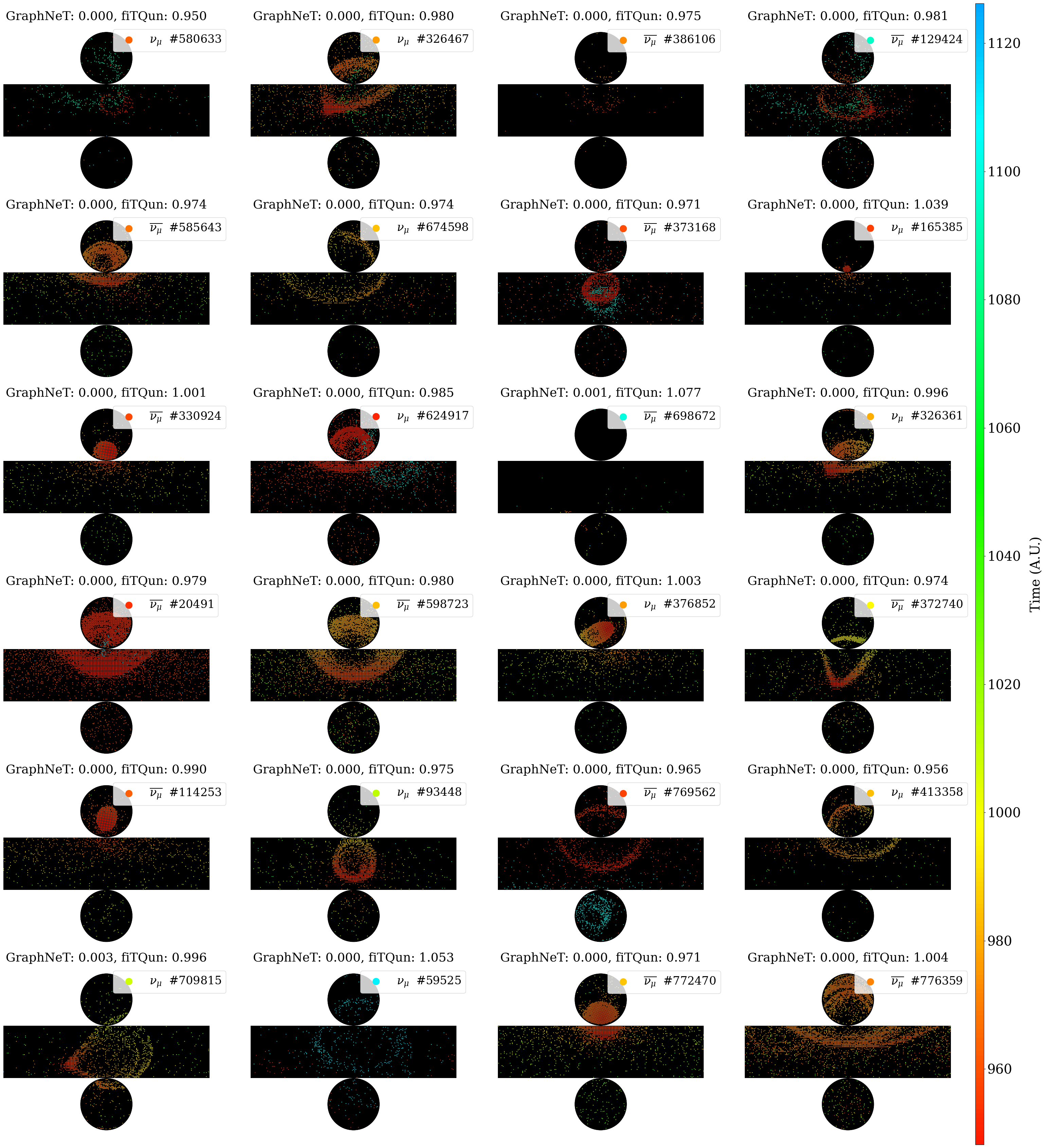}
    \caption{Events displays of randomly selected CC interaction neutrino events for which the GNN flavour classification method predicts correctly with great confidence, while the fiTQun-based method predicts incorrectly with great confidence. Several of the event displays show multiple rings of Cherenkov light, indicating that the GNN has the ability to accurately classify low energy muon neutrino events.}
    \label{gb_event_disp_fTime}
\end{figure*}

\newpage

\acknowledgments

Funded by the European Union. Views and opinions expressed are however those of the author(s) only and do not necessarily reflect those of the European Union. Neither the European Union nor the granting authority can be held responsible for them.

The authors acknowledge support provided by the European Union’s Horizon 2020 research and innovation programme under the Marie Skłodowska -Curie grant agreement No 860881-HIDDeN.

The training of models and reconstruction of events were performed using the GraphNeT framework (Apache 2.0), developed and maintained by GraphNeT group. The computations were enabled by resources provided by the National Academic Infrastructure for Supercomputing in Sweden (NAISS) and the Swedish National Infrastructure for Computing (SNIC) at LUNARC partially funded by the Swedish Research Council through grant agreements no. 2022-06725 and no. 2018-05973. 

The authors acknowledge the assistance from C. Vilela, E. O’Sullivan, H. Tanaka, B. Quilain and M. Wilking for the use of the WCSim and fiTQun software packages.






\bibliographystyle{JHEP}
\bibliography{biblio.bib}



\end{document}